\documentclass[11pt,a4paper]{article}

\usepackage[margin=2.5cm]{geometry}
\usepackage[utf8]{inputenc}
\usepackage[T1]{fontenc}
\usepackage[final,expansion=false,protrusion=true]{microtype}
\usepackage{amsmath,amssymb,amsfonts}
\usepackage{graphicx}
\usepackage{booktabs}
\usepackage{caption}
\usepackage{subcaption}
\usepackage{array}
\usepackage{hyperref}
\usepackage[numbers,sort&compress]{natbib}
\usepackage{float}

\hypersetup{colorlinks=true, linkcolor=blue, citecolor=blue, urlcolor=blue}
\newcommand{\Om}{\Omega_m}
\newcommand{\Ob}{\Omega_b}
\newcommand{\Or}{\Omega_r}
\newcommand{\Of}{\Omega_f}
\newcommand{\Ofz}{\Omega_{f,0}}
\newcommand{\rhoc}{\rho_{c,0}}

\begin{document}

\title{\large\bfseries The Umami Chaplygin Gas as a Unified Dark Fluid: Observational Constraints, Perturbative and Thermodynamic Stability Analysis}

\author{Trishit Banerjee\thanks{trishit23@gmail.com} \\[2pt]
\small Department of Mathematics, Netaji Subhash Engineering College,\\
\small Panchpota, Garia, Kolkata 700152, West Bengal, India
\and
Atreyee Biswas\thanks{atreyee11@gmail.com} \\[2pt]
\small Department of Applied Mathematics, Maulana Abul Kalam Azad University of Technology,\\
\small Haringhata, West Bengal 741249, India}

\date{}

\maketitle

\begin{abstract}
We study a cosmological scenario in which the Universe contains standard baryonic matter, radiation, and a single unified dark fluid, the umami Chaplygin gas, which accounts for both dark matter and dark energy with no separate dark-matter component.  We first analyse the thermodynamic behaviour of the fluid and show that, unlike earlier Chaplygin-type models, its thermodynamic stability is conditional rather than global: stability holds only above a critical density $\rho_1=\sqrt{|A/w|}$, the fluid reheats once this density is crossed, and a viable low-density regime requires $|w|<2$. Thermodynamic stability at the present epoch requires $|A|<|w|\rho_0^2$, a condition that is more easily met for $|w|>1$. We then constrain both the $|w|<1$ and $|w|>1$ cases with Type Ia supernovae, DESI DR2 baryon acoustic oscillations, cosmic-chronometer Hubble data and Planck 2018 distance priors through a Markov Chain Monte Carlo analysis. The $|w|<1$ case fits all four datasets well, with $w \simeq -0.73$, $\Omega_m^{\rm eff} \simeq 0.32$ and $H_0 \simeq 66.8$ km s$^{-1}$ Mpc$^{-1}$. It is statistically comparable to flat $\Lambda$CDM, with $\Delta{\rm AIC}=-4.8$ and $\Delta{\rm BIC}=+0.6$. Its best-fit parameters violate the stability condition. The $|w|>1$ case with a Gaussian prior satisfies the condition but fits worse, with $\Delta{\rm AIC}\simeq+66$ relative to $\Lambda$CDM. The best fit that obeys the condition exactly still has $\Delta{\rm AIC}\simeq+50$. Finally, a linear perturbation analysis shows that the adiabatic sound speed turns negative at exactly the same critical density, triggering exponential small-scale growth at late times in the observationally preferred case. These results reveal a tension between a good fit to background data and the physical stability of the umami unified dark fluid.
\end{abstract}


\section{Introduction}
\label{sec:intro}

Recent observations like Supernovae Ia (SNeIa) \cite{Riess1998,Perlmutter1999}, cosmic microwave background
(CMB)\cite{Bennet2003} , large scale structure (LSS) \cite{Hawkins2003,Tegmark2004,Cole2005,Eisenstein2005} etc suggest that our universe currently in an accelerating phase. One plausible way to describe this phenomenon is to assume the existence of an exotic matter called dark energy (DE) which has negative pressure. The most obvious choice for DE is the cosmological constant ($\Lambda$CDM model) \cite{Weinberg1989,Peebles2003} with constant equation of state
(EoS) $\omega=-1$. But it suffers from several drawbacks such as coincidence and fine tuning
problems \cite{Weinberg1989,Peebles2003}. As an alternative of cosmological constant, various fluid models have been proposed with variable EoS \cite{Caldwell1998} that can describe DE and has the potentiality of solving the problems persisting in $\Lambda$CDM model. Among these, Chaplygin gas type cosmology is very popular and promising one \cite{Bilic2002,Gorini2003}. This type of models behaves as dust in the early era while as DE in the later stage i.e unify DE with Dark matter and behaves as a single dark fluid and
therefore they belong to the class of unified model of dark energy. The original
Chaplygin gas model was proposed by Kamenshchik, Moschella, and Pasquier \cite{Kamenschick2001}
with EoS
\begin{eqnarray*}
P& =& -\frac{B}{\rho}
\end{eqnarray*}

Here $P$ and $\rho$ are the pressure and density of the fluid respectively and $B$ is a
constant. Later Bento et al.\cite{Bento2002} generalized the EoS to
\begin{eqnarray*}
P =-\frac{B}{\rho^{\alpha}}
\end{eqnarray*}
where $\alpha$($0\leq \alpha\leq 1$) generalizes from 1 to any arbitrary constant and this generalized model is called the generalized Chaplygin gas (GCG) model. The GCG model was tested with supernova and CMB data \cite{Makler2003,Bean2003}. Now, at high energy density, GCG model coincides with pressureless dust ($p=0$), which does not completely conform with our universe. Therefore, modified Chaplygin gas (MCG) \cite{Debnath2004} was introduced with the following equation of state
\begin{eqnarray*}
P = A\rho-\frac{B}{\rho^{\alpha}} 
\end{eqnarray*}
where $A$ is a positive constant. Unlike GCG, this model depicts constant negative pressure at low energy density and high pressure at high energy density- quite appropriate in describing different eras of evolution of universe.  $A = \frac{1}{3}$  is the best fitted value for MCG model \cite{Debnath2004} describing evolution of the universe from radiation regime to the Cold dark matter regime. Afterwards, many variations of this
model such as extended chaplygin (ECG)\cite{Pourhassan2014}, variable chaplygin gas \cite{Guo2007} have been introduced. H.B. Sandvik et al \cite{Sandvik2004} showed that this class of models faces a serious obstacle: the adiabatic sound speed associated with the equation of state is generically large enough to imprint unphysical oscillations on the matter power spectrum and the cosmic microwave background, and only survives observational scrutiny in the limit $\alpha\to0$, where the model becomes observationally indistinguishable from the standard $\Lambda$CDM cosmology. Recently a new generalization of this model namely Umami Chaplygin gas, has been proposed by R. Lazkoz et al \cite{Lazkoz2019} with EoS
\begin{equation}
P=-\frac{\rho}{\frac{1}{|w|}+\frac{\rho^2}{|A|}} \label{umami:eos}
\end{equation}
where $w$ and $A$ are real constants.
It was shown in \cite{Lazkoz2019}  Umami Chaplygin gas can not only unify dark energy and dark matter but also can unify dark fluid with baryonic matter, i.e it can contribute the total matter distribution of the universe. Therefore it is interesting to study this new generalization of Chaplygin gas model as an alternative of $\Lambda$CDM model. The \emph{umami} Chaplygin gas, introduced by R. Lazkoz \cite{Lazkoz2019}, modifies the early versions of Chaplygin type fluid construction so that the low-density limit approaches a generic constant-$w$ fluid rather than a pure cosmological constant. A phase-space (dynamical-systems) analysis of the background evolution of this fluid by Q. Huang et al \cite{Huang2021} identified regions of parameter space admitting a stable, late-time accelerating attractor. Several complementary approaches to the perturbative behaviour of Chaplygin-type unified fluids have been developed in the broader literature: a fully gauge-invariant relativistic treatment was given by V. Gorini et al \cite{Gorini2008}; non-adiabatic (entropy) perturbations that decouple the pressure response from the adiabatic sound speed were shown by R.~R.~R. Reis et al \cite{Reis2003} to remove the small-scale instability found in quartessence Chaplygin models; and A. Abdullah\cite{Abdullah2022} demonstrated that even where linear theory predicts runaway growth, a nonlinear spherical-collapse treatment can still permit structure formation above a model-dependent Jeans scale. More recently, M. Hashim, and A. El-Zant \cite{Hashim2025} implemented an explicit clustering prescription for a unified dark sector fluid confronted with current data while  P.~K.~S Dunsby, O. Luongo and M. Muccino \cite{Dunsby2024} proposed an alternative single-fluid unification with non-zero pressure. E. Frion et al \cite{Frion2024} examined whether unified dark-matter models with a background transition can relieve the Hubble tension. This tension is a mismatch between the early Universe and local values of $H_0$ \cite{DiValentino2021,Riess2022}. Recent baryon acoustic oscillation data from DESI \cite{DESI2024,DESI2025} also test late-time dark energy models. In this work we study the cosmological scenario in which the umami fluid is the sole constituent of the dark sector, so that the Universe is composed of
\begin{center}
\emph{baryonic matter} $+$ \emph{radiation} $+$ \emph{umami unified dark fluid (dark matter + dark energy)},
\end{center}
with no separate cold-dark-matter component. In this scenario, the fraction of the fluid that behaves as matter at any epoch is not an assumption but a direct consequence of the equation of state and the background solution it generates.
Also, in 2021, S. K. Biswas and A. Biswas \cite{Biswas2021} studied interacting Umami Chaplygin gas in flat FRW model of universe in context of thermodynamic and dynamical behaviour. In particular, considering Umami fluid as dark energy interacting with dark matter, they discussed irreversible thermodynamics both for apparent and event horizon as bounding horizon in two separate cases. Also,the authors found some cosmologically interesting critical points describing late time accelerated evolution of the universe attracted by cosmological constant and accelerated scaling attractor in quintessence era with some restrictions on model parameter $\omega$ and coupling parameter $\lambda$ to alleviate coincidence problem. At this point it is necessary to mention
that our universe behaves as an isolated thermodynamic system bounded by some
horizon \cite{Bekenstein1973,Hawking1975,GibbonsHawking1977}. The Friedmann equations follow from the first law of thermodynamics at the apparent horizon \cite{CaiKim2005,AkbarCai2007}. They also follow from the thermodynamics of spacetime \cite{Jacobson1995,Padmanabhan2010}. Though thermodynamic analysis can not predict any precise value of the observable parameters but can give limit of their values. Therefore, study of thermodynamic behaviour of the dark fluids which make up energy budget of the universe, is an useful tool to investigate different phenomenon during evolution of the universe.\\
The thermodynamics of dark energy fluids was also studied in \cite{Lima2004}. In a recent past FC. Santos et al studied thermodynamic stability of GCG \cite{Santos2006} and
MCG \cite{Santos2007} while D. Panigrahi investigated the same for VCG \cite{Panigrahi2015}. Also Chakraborty
et al in a series of works \cite{Chakraborty2017,Chakraborty2019,Chakraborty2017b} studied thermodynamic stability for different
types of dark fluids.

The paper is organised as follows. In Sec.~2 we describe the model: a spatially flat, homogeneous and isotropic universe containing baryons, radiation and the umami Chaplygin gas as the only dark component. We derive the background evolution of the fluid and reduce it to a one-dimensional root-finding problem, from which the full expansion history follows. In Sec.~3 we develop the thermodynamic description of the umami Chaplygin gas following Santos et al., analysing its pressure, equation-of-state and deceleration parameters, thermodynamic stability and thermal equation of state, and compare its behaviour with earlier Chaplygin-type models. This analysis identifies the parameter regions in which the fluid is thermodynamically stable at the present epoch and thereby motivates the two cases, $|w|<1$ and $|w|>1$, examined in the rest of the paper. In Sec.~4 we set out the observational analysis, describing the distance measures, the four datasets (Type Ia supernovae, DESI DR2 BAO, cosmic chronometers and Planck 2018 distance priors), the corresponding likelihoods and the MCMC procedure. Sec.~5 presents the resulting parameter constraints for both cases and compares their fits to the data and to flat $\Lambda$CDM. In Sec.~6 we examine the linear stability of density perturbations in the fluid, through its adiabatic sound speed and the coupled evolution of baryon and fluid density contrasts, and relate the resulting instability to the thermodynamic results of Sec.~3. Finally, in Sec.~7 we summarise our results and outline directions for future work.

\section{Model Description}\label{sec:model}

We consider our universe to be spatially flat, homogenous and isotropic whose line element is given by
\begin{equation}
ds^2 = -dt^2 + a(t)^2\left(dx^2+dy^2+dz^2\right),
\label{eq:metric}
\end{equation}
where $t$ is cosmic time and $a(t)$ is the scale factor, normalized to $a=1$ today. Spatial flatness is adopted throughout this work, consistent with current observational constraints. Inserting Eq.~\ref{eq:metric} into Einstein's field equations, sourced by a mixture of perfect fluids with total energy density $\rho_{\rm tot}$ and total pressure $p_{\rm tot}$, yields the two independent Friedmann equations:
\begin{eqnarray}
  H^2 \equiv \left(\frac{\dot{a}}{a}\right)^2 & = & \frac{8\pi G}{3}\rho_{tot} \label{eq:friedmann1}\\
\frac{\ddot {a}}{a} &= & -\frac{4\pi G}{3}\left(\rho_{tot}+3p_{tot}\right) \label{eq:friedmann2} \\
 \end{eqnarray}

where $H$ is the Hubble parameter. The continuity equation for the total cosmic fluid reads as
\begin{equation}
\dot\rho_{\rm tot} + 3H\left(\rho_{\rm tot}+p_{\rm tot}\right) = 0.
\label{eq:continuity_total}
\end{equation}

We take the total cosmic fluid to be a mixture of three non-interacting components -- baryonic matter, radiation, and the umami dark fluid -- so that $\rho_{\rm tot}=\rho_b+\rho_r+\rho_f$ and $p_{\rm tot}=p_b+p_r+p_f$. Because the three components exchange energy with one another only gravitationally, through their joint appearance in Eq.~\ref{eq:friedmann1}, and not directly, each component separately satisfies its own continuity equation of the same form as Eq.~\ref{eq:continuity_total},
\begin{equation}
\dot\rho_i + 3H\left(\rho_i+p_i\right) = 0, \qquad i=b,r,f.
\label{eq:continuity_i}
\end{equation}
For baryons, $p_b=0$, and Eq.~\ref{eq:continuity_i} integrates immediately to $\rho_b\propto a^{-3}$, or in terms of the density parameter $\Omega_b(a)\equiv\rho_b(a)/\rhoc$,
\begin{equation}
\Omega_b(a) = \Omega_{0b}a^{-3},
\label{eq:omegab}
\end{equation}
where $\Omega_{0b}$ is the current value of baryon density parameter. For radiation, $p_r=\rho_r/3$, and the same integration gives
\begin{equation}
\Omega_r(a) = \Omega_{0r}a^{-4},
\label{eq:omegar}
\end{equation}
with $\Omega_{0r}\equiv\Omega_r(a=1)=9.24\times10^{-5}$ fixed to the standard photon-plus-neutrino value. Both results are the familiar scalings of pressureless matter and radiation, obtained here directly from Eq.~\ref{eq:continuity_i} as a special case.

The umami fluid does not admit such a simple scaling, because it is not a fluid of constant or power-law equation of state. It is instead defined by the non-linear equation of state given by eqn. (\ref{umami:eos}).
From here on, $\rho\equiv\rho_f$ and $P\equiv p_f$ denote the density and pressure of this fluid alone. Umami fluid interpolates smoothly between two physically distinct regimes,
\begin{align}
\rho \to \infty: \qquad & P \;\longrightarrow\; -\frac{|A|}{\rho}, \label{eq:eos_highrho}\\[4pt]
\rho \to 0: \qquad & P \;\longrightarrow\; -|w|\,\rho. \label{eq:eos_lowrho}
\end{align}
Equation~\ref{eq:eos_highrho} is the equation of state of the original Chaplygin gas, so that at high density -- early cosmic times -- the umami fluid behaves as pressureless matter. Equation~\ref{eq:eos_lowrho}, by contrast, is the equation of state of a barotropic fluid with constant equation-of-state parameter $w$, rather than the pure cosmological constant obtained in the low-density limit of the original Chaplygin gas. It is this single equation of state, with no further assumption or additional component, that allows the fluid to play the role of dark matter at early times and of dark energy at late times.

Substituting $P_f$ from eq.(\ref{umami:eos}) into the fluid's own continuity equation (\ref{eq:continuity_i}) with $i=f$, we obtain an implicit form of $\rho(a)$ as follows:
\begin{align}
&\rho^{\frac{1}{1-|w|}}\Big[\,|A|-|A||w|+|w|\rho^{2}\,\Big]^{\frac{|w|}{2(|w|-1)}}
\notag\\
&\;=
\rho_{0f}^{\frac{1}{1-|w|}}\Big[\,|A|-|A||w|+|w|\rho_{0f}^{2}\,\Big]^{\frac{|w|}{2(|w|-1)}}\,a^{-3},
\label{eq:implicit_dimful}
\end{align}
where the subscript $0$ again denotes evaluation today. Introducing the dimensionless density $\Of\equiv\rho/\rhoc$ and the correspondingly rescaled parameter $A\equiv A/\rhoc^2$ turns eq.~\ref{eq:implicit_dimful} into
\begin{align}
&\Of^{\frac{1}{1-|w|}}\Big[\,|A|-|A||w|+|w|\Of^{2}\,\Big]^{\frac{|w|}{2(|w|-1)}}
\notag\\
&\;=
\Ofz^{\frac{1}{1-|w|}}\Big[\,|A|-|A||w|+|w|\Ofz^{2}\,\Big]^{\frac{|w|}{2(|w|-1)}}\,a^{-3}.
\label{eq:implicit_dimless}
\end{align}
Both sides of Eq.~\ref{eq:implicit_dimless} have the same functional form, evaluated at $\Of(a)$ on the left and at $\Ofz$ on the right; it is therefore convenient to define this common function once and for all,
\begin{equation}
F(x) \equiv x^{\frac{1}{1-|w|}}\Big[\,|A|(1-|w|)+|w|\,x^{2}\,\Big]^{\frac{|w|}{2(|w|-1)}},
\label{eq:Fdef}
\end{equation}
so that Eq.~\ref{eq:implicit_dimless} takes the compact form used throughout the remainder of this paper,
\begin{equation}
F(\Of(a)) = F(\Ofz)\,a^{-3}.
\label{eq:implicit}
\end{equation}
Unlike Eqs.~\ref{eq:omegab}--\ref{eq:omegar}, Eq.~\ref{eq:implicit} cannot be inverted algebraically to give $\Of(a)$ directly. It can, however, be shown that the function $F$ defined in Eq.~\ref{eq:Fdef} is monotonically increasing on $(0,\infty)$: as $x\to0^+$, $F(x)\to0$, while as $x\to\infty$, $F(x)\sim x$, since the two exponents in Eq.~\ref{eq:Fdef} sum to unity. A monotonic function takes every value in its range exactly once, so for every scale factor $a$ -- for which the right-hand side of Eq.~\ref{eq:implicit} is simply a known number, once $\Ofz$ is fixed by flatness -- there is exactly one value of $\Of(a)$ satisfying it. Finding $\Of(a)$ is therefore reduced to a one-dimensional root-finding problem: for each $a$ of interest, we search numerically for the value of $x$ at which $F(x)$ equals the known right-hand side, using a bisection-type algorithm that is guaranteed to converge precisely because $F$ is monotonic. Repeating this at a set of scale factors spanning the range needed for the observational analysis, and interpolating smoothly between them with a cubic spline in $(\ln a,\ln\Of)$, yields a continuous function $\Of(a)$ that can be evaluated at any redshift required by the likelihood of Sec.~\ref{sec:obs}.

Having obtained $\Of(a)$, $\Omega_b(a)$, and $\Omega_r(a)$ separately, we now return to the total Friedmann equation, Eq.~\ref{eq:friedmann1}, and assemble the three components. Dividing through by $H_0^2$ and writing the result in terms of the dimensionless Hubble function $E(z)\equiv H(z)/H_0$, with $1+z=1/a$,
\begin{align}
E(z) &= \sqrt{\Ob(1+z)^{3}+\Or(1+z)^{4}+\Of(z)}, \label{eq:friedmann}\\
\Ofz &= 1-\Ob-\Or,
\notag
\end{align}
where the relation for $\Ofz$ follows from evaluating Eq.~\ref{eq:friedmann} at $z=0$ and imposing $E(0)=1$, i.e.\ spatial flatness.

Equation~\ref{eq:friedmann} is sufficient to compute the full background expansion history. As anticipated at the start of this section, however, the observational analysis of Sec.~\ref{sec:obs} additionally requires an estimate of the total matter content of the Universe at early times, for use in fitting formulas that were originally calibrated assuming a standard matter-plus-radiation early Universe. Rather than introducing an independent parameter for this purpose, we obtain it directly from the fluid's own solution. Because the umami fluid behaves as dust at high density, the monotonicity of $F$ established above implies that $\Of(a)\to\kappa\,a^{-3}$ as $a\to0$, where the coefficient
\begin{equation}
\kappa \equiv \lim_{a\to0}\Of(a)\,a^{3}
\label{eq:kappa}
\end{equation}
is fixed entirely by $(w,A,\Ofz)$ through Eq.~\ref{eq:implicit}, and is therefore not an independent parameter of the model but a derived property of the background solution already in hand. We use it to define the effective total-matter density parameter
\begin{equation}
\Om^{\rm eff} \equiv \Ob+\kappa,
\label{eq:Omeff}
\end{equation}
which plays the role of the matter density parameter wherever a standard, $\Lambda$CDM-calibrated fitting formula for the early Universe is required, without altering the genuine background expansion rate of Eq.~\ref{eq:friedmann}. The background cosmology developed in this section is fully specified by four free parameters: the baryon density $\Ob$, the reduced Hubble constant $h$ (with $H_0=100h$~km~s$^{-1}$~Mpc$^{-1}$), and the two equation-of-state parameters $w$ and $\tilde A\equiv-\log_{10}A$.
\section{Thermodynamic Behaviour Analysis}
\label{sec:thermo}
To study its thermodynamic behaviour, we regard the umami Chaplygin gas as a one-component fluid whose particle number $N$ remains constant during expansion. Its internal energy $U$ and pressure $P$ can then be expressed as functions of the two natural variables, entropy $S$ and volume $V$. The pressure of a fluid with internal energy U confined within a volume V can be evaluated as \cite{Callen1985}
\begin{equation}
\left(\frac{\partial U}{\partial V}\right)_S=-P\label{3}
\end{equation}

Also, the uniform density of the fluid is defined by
\begin{equation}
\rho=\frac{U}{V}\label{4}
\end{equation}
Combining Eqs.(\ref{umami:eos}), (\ref{3}) and (\ref{4}) one can obtain that
\begin{equation}
\left(\frac{\partial U}{\partial V}\right)_S=\frac{\frac{U}{V}}{\frac{1}{|w|}+\frac{\frac{U^2}{V^2}}{|A|}}\label{5}
\end{equation}

\begin{figure}
\includegraphics[width=0.6\textwidth]{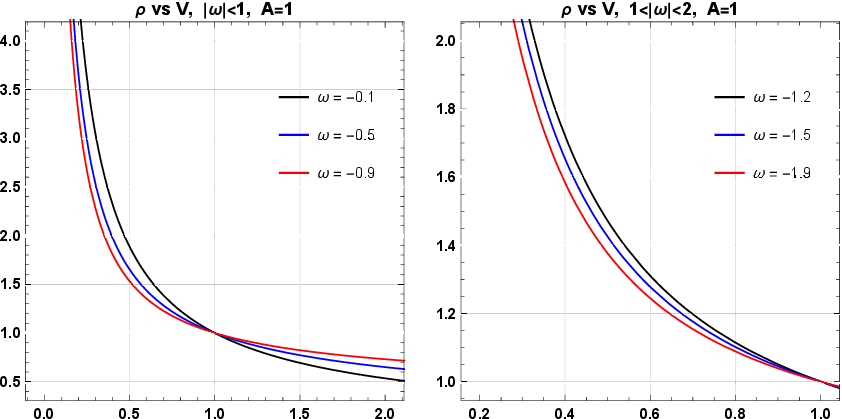}
\caption{Density $\rho$ against $V$ for $|\omega|<1$ and $|\omega|>1$ with $A=1,c=1$.}
\label{fig:1}
\end{figure}

If we integrate eqs.($\ref{5}$), from resulting expression it will not be possible to express U
explicitly in terms of V and the constant of integration. Rather we obtain the following  equation for $V$ in terms of energy density $\rho$:
\begin{equation}
V^{2\alpha}=c\rho^{-2}\left(\alpha+\rho^2|\frac{\omega}{A}|\right)^{|\omega|}\label{6}
\end{equation}
where $\alpha=1-|\omega|$ and $c>0$ is a constant of integration and which may be either universal constant or a function of entropy only i.e, $c = c(S)$. 
Now,from equation (\ref{6}) one can write 
\begin{equation}
\frac{c^{\frac{1}{2\alpha}}}{V}=\frac{\rho^{\frac{1}{\alpha}}}{\left(\frac{\alpha}{\rho^2}+|\frac{\omega}{A}|\right)^{\frac{|\omega|}{2\alpha}}}\label{7}
\end{equation}
When $V\ll c^{\frac{1}{2\alpha}}$, $\rho$ becomes infinitely large i.e $\rho\rightarrow \infty $. On the other hand for large volume, when $V\gg c^{\frac{1}{2\alpha}}$, then  $\rho\rightarrow 0$ for $\alpha>0$ and $\rho\rightarrow \sqrt{-\alpha|\frac{A}{\omega}|}$ for $\alpha<0$. Infact, if $\alpha<0$, $V$ will take real value only for $\rho>\sqrt{|A|\left(1-\frac{1}{|\omega|}\right)}$. So, for $\alpha<0$, the minimum value $\rho$ can take is $\sqrt{|A|\left(1-\frac{1}{|\omega|}\right)}=\rho_{min}$(say). This behaviour of $\rho$ with the change of volume $V$ for both $|\omega|<1$ and $|\omega|>1$ can clearly be comprehended from fig-\ref{fig:1}. 
We observe that in the initial period when $V\approx 0$, $\rho$ starts from a very high value and after that the $\rho$ drastically falls down close to zero within a short range of $V$ and after that as $V $ increases it gradually monotonically decreases towards zero for $\alpha>0$ or towards $\rho_{min}$ for $\alpha<0$. 
This scenario is compatible with the observational results.
\subsection{\textbf{Pressure}}

\begin{figure}
\includegraphics[width=0.6\textwidth]{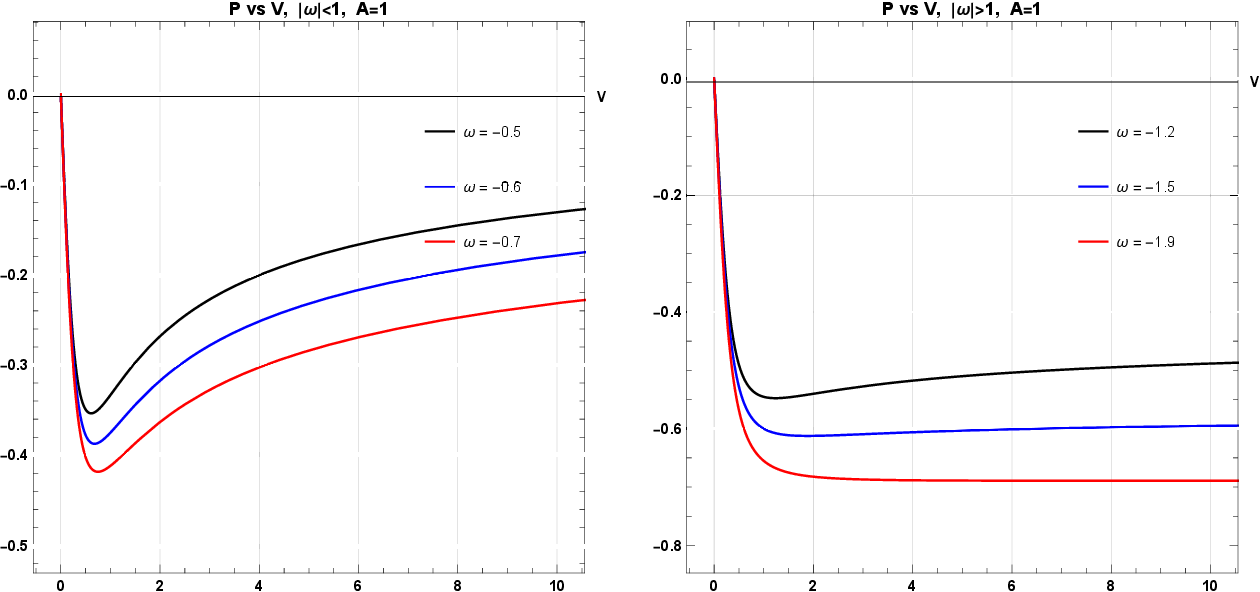}
\caption{Pressure $P$ against $V$ for $|\omega|<1$ and $|\omega|>1$ with $A=1,c=1$.}

\label{fig:2}
\end{figure}

The pressure $P$ of the Umami fluid can not be explicitly expressed in terms of volume $V$. But from equation (\ref{7}) we can infer that for small volume the pressure of Umami fluid behaves as $P\approx -\frac{|A|}{\rho}$. So for sufficiently large $\rho$, $P\approx 0$. On the other hand, for large volume   $P\approx -|\omega|\rho$, i,e it behaves as a constant Equation of State (EoS) dark energy. However, from the expression of $P$ in equation (\ref{umami:eos}) it is evident that for finite $\rho$ Umami Chaplygin gas always has negative pressure while for sufficiently large $\rho(\rho\rightarrow\infty)$ it behaves as dust just like other Chaplygin gases. So Umami fluid perfectly behave as an unified dark matter. To understand the behaviour of pressure $P$ with the change of $V$ we plotted $P$ against volume $V$ in figure-\ref{fig:2}. 
 
From this figure, we observe that, all the $P-V$ curves start at $P=0,V=0$. After that in case of $|\omega|<1$, within a small range of $V$, they plunge almost vertically to a minimum and smaller value of $\omega$ lowers this minimum value. The steep downward dive is then followed by a monotonic increase of $P$ but $P$ always remains negative thereafter. On the other hand, for $|\omega|>1$, pressure monotonically decreases with negative value with the increase of volume after it crosses the point $V=0$. So, at the initial point when $V= 0$, Umami fluid starts as pressureless dark matter and thereafter for $V>0$ it becomes negative but nature of evolution of pressure across $V$ are very different from each other for  $|\omega|<1$ and $|\omega|>1$.

\subsection{\textbf{EoS and Deceleration Parameter}}
The equation of state (EoS) parameter $W$ and deceleration parameter $q$  are defined as
\begin{eqnarray*}
  W &=&\frac{P}{\rho}\\
  q &=& \frac{1}{2}+\frac{3}{2}W
\end{eqnarray*}

Both are again implicit relation with $V$ in view of equation (\ref{6}). From the expression of $W$ we observe since $P=0$  and $\rho\rightarrow \infty$ at $V=0$, $W$ starts right from zero which can be attributed to pressureless dark matter. For large volume  $W\approx -|w|$. In quintessence era $-1<W<-\frac{1}{3}$ which imposes restriction on the parameter $w$ : $\frac{1}{3\left(1-\frac{\rho_0^2}{3|A|}\right)}<|w|<\frac{1}{\left(1-\frac{\rho_0^2}{|A|}\right)}$ while for $|\omega|>\frac{1}{\left(1-\frac{\rho_0^2}{|A|}\right)}$ it corresponds to phantom era. $\rho_0$ denotes here the present value of $\rho$ .
  \begin{figure}

\includegraphics[width=0.6\textwidth]{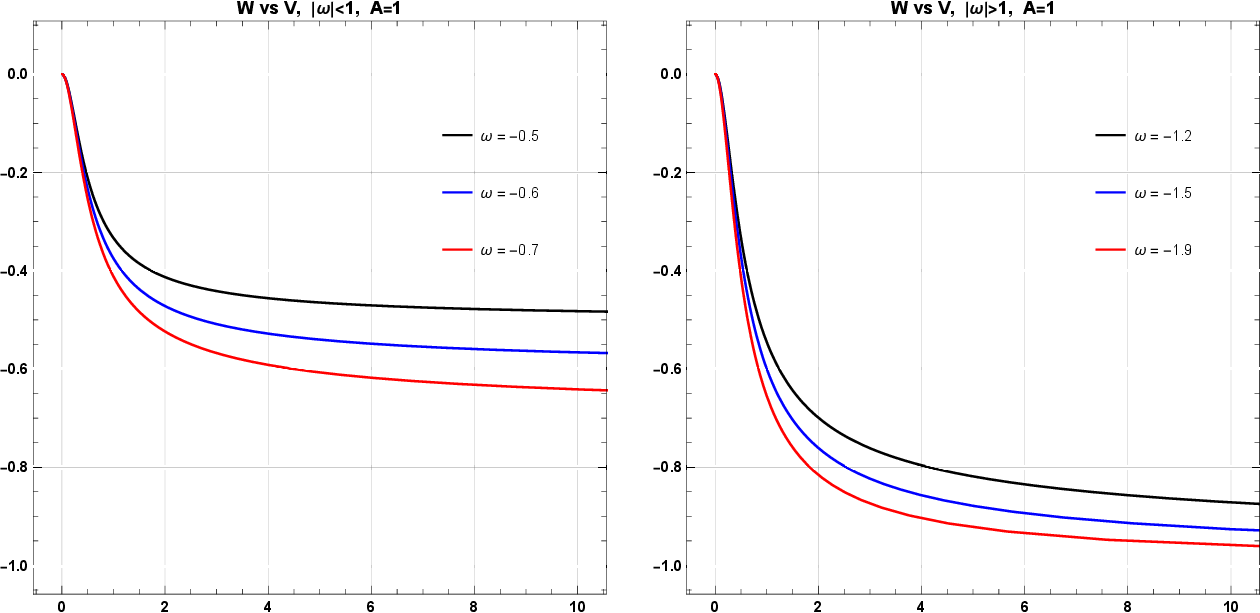}
\caption{Effective equation-of-state parameter $W=P/\rho$ against $V$ for $|\omega|<1$ and $|\omega|>1$ with $A=1,c=1$ . Both branches start at $W=0$ and fall toward a constant value, approaching standard dark energy behaviour at large $V$.}
\label{fig:3}
\end{figure}

\begin{figure}
\includegraphics[width=0.6\textwidth]{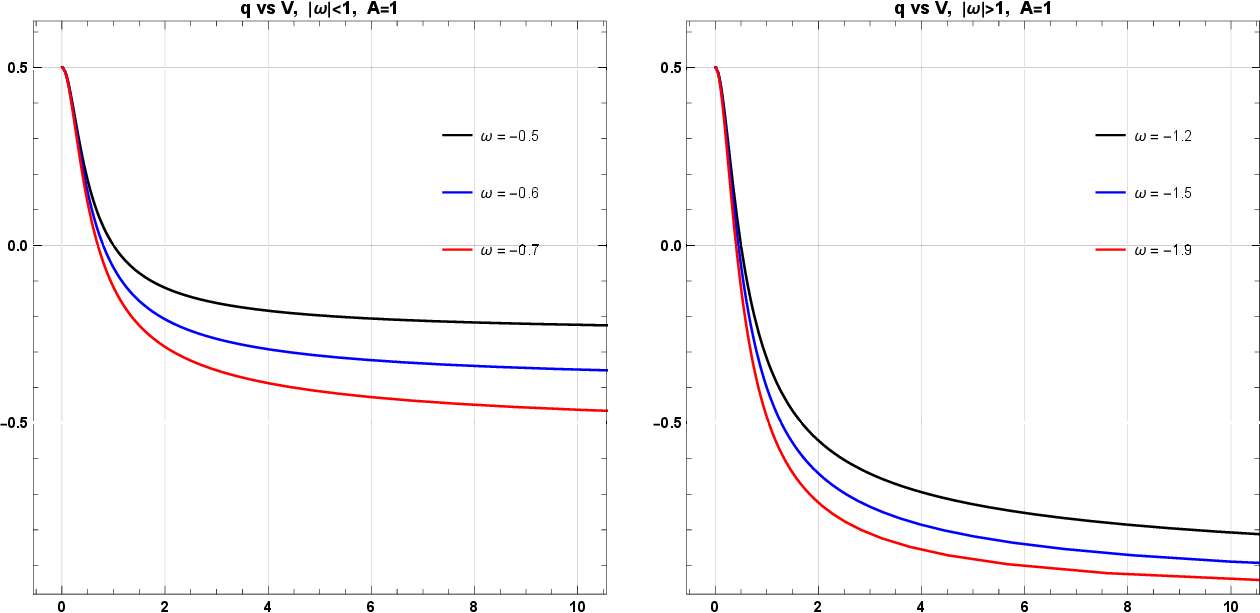}
\caption{Deceleration parameter $q$ against $V$ for $|\omega|<1$ and $|\omega|>1$ with $A=1,c=1$.}
\label{fig:4}
\end{figure}

In Figure-\ref{fig:3} we have shown the evolution curve of $W$ with the expansion of $V$. As expected we see $W=0$ at $V=0$ and then $W$ curve monotonically decreases remaining negative through out. However, for the universe to be accelerating, it is essential $W$ to be less than $-1/3$. For $-1/3\leq W \leq 0$ the universe decelerates. From the $W-V$ curve we see that there exists a certain value of $V$, say $V=V_{acc}=\left(\frac{c(2|\omega|)^{2|\omega|}}{\left(3|\omega|-1)|\frac{A}{\omega}|\right)}\right)^{1/2\alpha}$ such that for $0\leq V\leq V_{acc}$, we have $-1/3\leq W \leq 0$. 
Therefore from this plot it is apparently evident that Umami Chaplygin fluid can behave as dark matter with zero-pressure at the starting time and with negative non-zero pressure for $0<V\leq V_{acc}$ . After that for the entire range of $V>V_{acc}$ it potentially behaves as dark energy. The numerical value of $V_{acc}$ increases with higher value of $|\omega|$. This behaviour can more clearly be comprehended from the deceleration curve in fig-\ref{fig:4}. But it should be noted that for $V=0$, the energy density $\rho\rightarrow \infty$ indicating singularity. Therefore we should restrict our study only for $V> 0$.\\

In the plot(fig-\ref{fig:4}) for $q$ against $V$, it is clearly seen that $q>0$ or $<0$ according as $V<V_{acc}$ or $V\ge V_{acc}$.  It is noteworthy, for $|\omega|<1$, $V_{acc}$ is very close to the starting point, i.e decelerating phase of universe is very short and restricted within a very short range from starting point $V=0$. On the contrary, in $|\omega|>1$ case, we observe that $q-V$ curve monotonically decreases from positive to negative and $V_{acc}$ depending  on the value of $\omega$ occurs at decent distance from starting point indicating longer decelerating phase of universe as expected. For $|\omega|<1$ after $q-V$ curve crosses $V_{acc}$, it becomes flat i.e Umami fluid becomes constant $W$ dark energy and the curves for different $\omega$ never meet each other. On the other hand for $|\omega|>1$, all the curves irrespective of the value of $\omega$ approaches $q=-1$, i.e the De-Sitter value. So the $|\omega|>1$ case is the one that resembles $\Lambda$CDM: a long matter-dominated phase, a smooth passage through $q=0$, and a de Sitter future. The $|\omega|< 1$ case is closer to a quintessence model with fixed $W$.

\subsection{\textbf{Stability Analysis}}
The thermodynamic stability of the system requires to satisfy the following necessary conditions:\\
\begin{itemize}
  \item (i) $\left(\frac{\partial P}{\partial V}\right)_{S}<0$, $\left(\frac{\partial P}{\partial V}\right)_{T}\leq 0$.
  \item (ii) thermal capacity at constant volume should be positive i.e $C_{V}>0$
\end{itemize}
Using Eqs. (\ref{6}) we calculate $\left(\frac{\partial P}{\partial V}\right)_{S}$ as follows:
\begin{eqnarray}
\left(\frac{\partial P}{\partial V}\right)_{S}&=&\frac{\frac{\partial P}{\partial \rho}}{\left(\frac{\partial V}{\partial \rho}\right)_{S}}\nonumber\\
~&=& -\frac{|\omega|\rho\left(|\frac{\omega}{A}|\rho^2-1\right)\left(\alpha+|\frac{\omega}{A}|\rho^2\right)}{V\left(1+|\frac{\omega}{A}|\rho^2\right)^3}\label{13}
\end{eqnarray}
 Now,  $\rho=\sqrt{|\frac{A}{\omega}|}=\rho_1$ when $V=\left\{c|\frac{\omega}{A}|(1+\alpha)^{|\omega|}\right\}^{1/2\alpha}=V_1$(say). We discuss negativity condition of $\left(\frac{\partial P}{\partial V}\right)_{S}$ in two separate cases.\\
\paragraph{\textbf{ Case-1}:\underline{\textbf{ $\alpha>0$ }}}
~\\
 
 it is inferred from equation (\ref{13}) that $\left(\frac{\partial P}{\partial V}\right)_{S}<0$ for $V< V_1$, i.e when $\rho>\rho_1$.
   At $V=V_1$ as expected  $\left(\frac{\partial P}{\partial V}\right)_{S}$ becomes zero and after $V$ crosses $V_1$, it remains positive until $V$ becomes sufficiently larger than $c^{\frac{1}{2\alpha}}$. When $V\gg c^{\frac{1}{2\alpha}}$ , $\rho\rightarrow 0$ and consequently $\left(\frac{\partial P}{\partial V}\right)_{S}$ also converges to zero.
  
\begin{figure}
\includegraphics[width=0.6\textwidth]{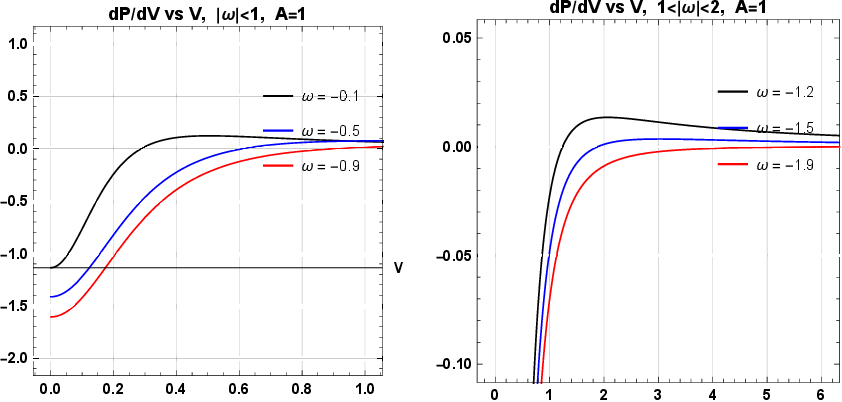}
\caption{$T$ against $V$ for $|\omega|<1$ branch and $1<|\omega|<2$ branch with $A=1, c=1$. }

\label{fig:5}
\end{figure}

 \paragraph{\textbf{ Case-2}:\underline{\textbf{ $\alpha<0$ }}}
 ~\\
  If $\alpha<0$, as already mentioned, the minimum value that $\rho$ can take is $\sqrt{|A|\left(1-\frac{1}{|\omega|}\right)}=\sqrt{|\alpha||\frac{A}{\omega}|}=\rho_{min}$ in order to keep $V$ real valued and $\rho$ attains this minimum value only when $V$ takes very large value. Now, since $\rho_{min}=\sqrt{|\alpha||\frac{A}{\omega}|}$ is minimum value of $\rho$, we must have $\sqrt{|\alpha||\frac{A}{\omega}|}<\sqrt{|\frac{A}{\omega}|}$ i.e $|\alpha|<1$. It means that $|\omega|$ will must lie between 1 and 2 for $\alpha<0$ case. Therefore from eqn.(\ref{13}) again it can be seen that for negative $\alpha$ also  $\left(\frac{\partial P}{\partial V}\right)_{S}<0$ when $V<V_1$. However, higher value of $|\omega|$ increases the numerical value of $V_1$ which for $|\omega|\approx 2$ tends to $\infty$, i.e then  $\left(\frac{\partial P}{\partial V}\right)_{S}<0$ holds for the entire valid region of $V$. This behaviour of $\left(\frac{\partial P}{\partial V}\right)_{S}$ can clearly be understood from figure-\ref{fig:5}.

\begin{figure}
\includegraphics[width=0.6\textwidth]{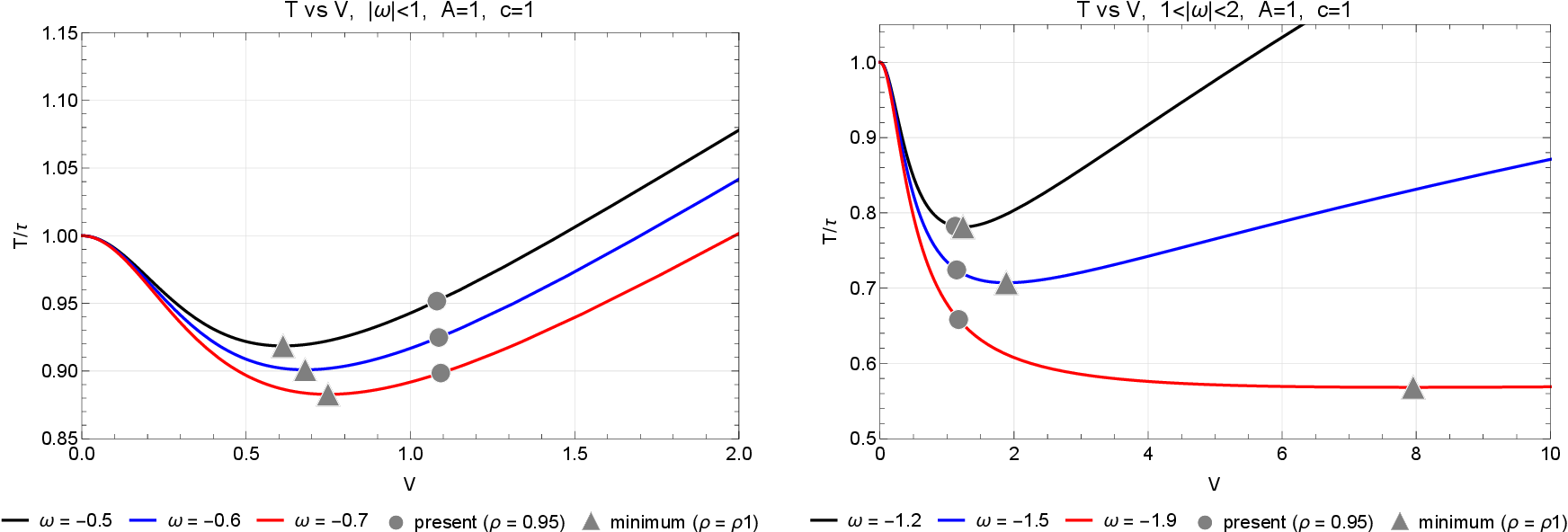}
\caption{$T$ against $V$ for $|\omega|<1$  and $1<|\omega|<2$ branch with $A=1,c=1$. }
\label{fig:6}
\end{figure}

In order to check whether $\left(\frac{\partial P}{\partial V}\right)_{T}<0$ , we have to express pressure $P$ in terms of Volume $V$ and temperature $T$, i.e $P=P(V,T)$.\\
Now, temperature $T$ can be calculated from the following relation:
\begin{equation}
  T=\frac{\partial U}{\partial S}\label{15}
\end{equation}
Using Eqn.(\ref{6}) and (\ref{15}) we then calculate $T$ as follows:
\begin{equation}
T=\frac{U \frac{\partial c}{\partial S}\left(\alpha+|\frac{\omega}{A}|\rho^2\right)}{2c\alpha \left(1+|\frac{\omega}{A}|\rho^2\right)}\label{16}
\end{equation}
Here $c$ is either an universal constant or function of $S$ only. If $c$ is assumed to be universal constant then $\frac{\partial c}{\partial S}=0$ and this implies that temperature $T$ will remain zero for all values of $V$. But we need temperature to be varied during expansion of universe, so $\frac{\partial c}{\partial S}\neq 0$.
From dimensional analysis we observe from eqn.(\ref{6}) that
\begin{equation}
[U]=[T][S]=\left|\frac{\omega}{A}\right|^{\frac{|\omega|}{2\alpha}}[c]^{\frac{1}{2\alpha}}\label{17}
\end{equation}
which gives
\begin{equation}
c=(\tau S)^{2\alpha}\left|\frac{A}{\omega}\right|^{|\omega|}\label{18}
\end{equation}
where $\tau$ has dimension of temperature .
Therefore we have
\begin{equation}
\frac{\partial c}{\partial S}=\frac{2\alpha c}{S}\label{19}
\end{equation}
Using Eqs.(\ref{16}) and (\ref{19}) we calculate the final expression for $T$ as follows:
\begin{equation}
T=\tau \left(\frac{\rho_1}{\rho}\right)^{\frac{|\omega|}{\alpha}}\frac{ \left(\alpha+|\frac{\omega}{A}|\rho^2\right)^{1+\frac{|\omega|}{2\alpha}}}{\left(1+|\frac{\omega}{A}|\rho^2\right)}\label{20}
\end{equation}
It is observed from eqn.(\ref{20}) that $\rho$ depends only on $T$, not on $V$ and consequently $P$ also becomes a function of $T$ only. It implies  that  $\left(\frac{\partial P}{\partial V}\right)_{T}=0$. This satisfies the stability criterion as defined by Santos et al\cite{Santos2006,Santos2007}.\\
Also, we observe that for the  limiting case when $\rho\rightarrow\infty$ at initial time, $T$ converges to the finite value $\tau$ for both $\alpha>0$ and $\alpha<0$. The lower viable limiting value of $\rho$ are although different for $\alpha>0$ and $\alpha<0$, $T$ diverges to $\infty$ when $\rho$ approaches it's respective minimum values for $\alpha>0$ and $\alpha<0$ case . However, the minimum value of $T$ is attained at $\rho=\rho_1$ and $T_{min}=\frac{\tau}{2}(1+\alpha)^{\frac{|\omega|}{2}+1}$. Since,$|\alpha|<1$, it is obvious that $T_{min}>0$. Therefore $0<\frac{\tau}{2}(1+\alpha)^{\frac{|\omega|}{2}+1}<\tau<\infty$. Also, it is to be noted that $T$ attains it's minimum value not when $\rho\rightarrow 0$, rather at $\rho=\rho_1$ gradually reducing from its limiting value $\tau$ and after it crosses $\rho_1$ at $V=V_1$, $T$ again starts increasing gradually. In figure-\ref{fig:6} evolution of temperature $T$ across the volume $V$ for different $\omega$ is shown. Since the unified fluid replaces both dark matter and dark energy, the flatness condition
$\Omega_{f,0}+\Omega_{b,0}+\Omega_{r,0}=1$, which follows from the Friedmann equation independently of the equation of state~\cite{Weinberg2008,Copeland2006}, gives $\rho_0\equiv\rho_{f,0}/\rho_{c,0}=1-\Omega_{b,0}-\Omega_{r,0}\simeq0.95$, with
$\Omega_{b,0}\simeq0.05$~\cite{Planck2018} and $\Omega_{r,0}\sim10^{-4}$~\cite{Fixsen2009}.

  Interestingly we observe from this figure that present value of $T$($=T_0$) can occur before or after $T_{min}$ depending on the value of $\omega$ and $A$. The precise condition for $T_0$ to occur before $T_{min}$ occurs is that the relation $\rho_0^2>|\frac{A}{\omega}|$ should be satisfied . So, a suitable adjustment between $\omega$ and $A$ can result in a stable present regime. Taking $|\omega|>1$ gives $A$ more flexibility to take wider range of values. In our figure we have fixed $A=1$ and accordingly observe that for $|\omega|<1$, $T_0$ always occurs after $T_{min}$ i.e $V_0> V_1$ so that for this case we are presently on the rising branch of $T$ and consequently the present time falls in the unstable regime. On the other hand, for $|\omega|>1$, the present epoch lies at or before this minimum, so that the fluid is in cooling period and thermodynamically stable throughout its past evolution. So, the temperature evolves differently in UCG case unlike GCG and MCG. For GCG/ MCG, the fluid cools monotonically during expansion, with no critical point or phase-transition-like behaviour.  The late reheating branch is new to UCG, and it coincides exactly with the unstable regime. In the next section after we find best fit value of $A$ and $\omega$ through observational analysis, the exact scenario can be understood, i.e it can be checked whether at present we are actually in stable or unstable regime.
Now, thermal capacity at constant volume is defined as $C_{V}=T\left(\frac{\partial S}{\partial T}\right)_V=V\left(\frac{\partial \rho}{\partial T}\right)_V$. Using (\ref{20})  we calculate $C_{V}$ as follows:
\begin{equation}
C_V=\frac{\rho V\left(\alpha+|\frac{\omega}{A}|\rho^2\right)\left(1+|\frac{\omega}{A}|\rho^2\right)}{|\omega| T\left(|\frac{\omega}{A}|\rho^2-1\right)}\label{25}
\end{equation}
Since $T$ and $V$ both are positive quantity, $C_V>0$ for $\rho>\rho_1$ which is same as obtained from the other stability conditions.
Now, we shall examine the well known thermodynamical relation 
\begin{equation}
\left(\frac{\partial U}{\partial V}\right)_T=T\left(\frac{\partial P}{\partial T}\right)_V-P\label{27}
\end{equation}
Using equations (\ref{umami:eos}), (\ref{6}) and (\ref{20}) we finally verify that
\begin{eqnarray}
\left(\frac{\partial U}{\partial V}\right)_T&=\rho =T\left(\frac{\partial P}{\partial T}\right)_V-P\label{28}
\end{eqnarray}
The Umami Chaplygin Gas (UCG) differs from earlier versions of Chaplygin gas (GCG and MCG) mainly by the fact that thermodynamic Stability is global for GCG/MCG but conditional for UCG. Santos et al \cite{Santos2006,Santos2007}. found the generalized Chaplygin gas (GCG) satisfies all the stability conditions throughout the whole expansion. They reached a similar conclusion for the modified Chaplygin gas (MCG). But for UCG, thermodynamical stability holds only for $\rho>\rho_1$. The earlier models have no comparable critical density. 

\subsection{Thermal Equation of State}

The analysis so far has described the umami Chaplygin gas through its density, pressure and volume, with the temperature obtained as a derived quantity in Eq.~\eqref{20}. For a complete thermodynamic description, however, a fluid must also be characterised by its thermal equation of state, which relates its pressure and internal energy to the temperature $T$ and volume $V$. Such a relation is the counterpart, for an exotic cosmological fluid, of the ideal-gas law $PV = NkT$ for an ordinary gas, and it allows the behaviour of the fluid to be followed directly as the Universe heats or cools during expansion. It also provides an independent check on the results of the previous subsections, since the limiting forms of $P(T)$ and $W(T)$ must reproduce the dust-like behaviour at early times and the dark-energy behaviour at late times found earlier. To obtain this relation, the thermodynamic quantities must be expressed in terms of $T$ and $V$. From Eq.~(\ref{20}), however, $\rho$ depends on $T$ alone, though only implicitly.
 Consequently we can derive an implicit form of internal energy $U=U(V,T)$ as follows :
\begin{equation}\label{ U(T)}
  U=V\rho(T)
\end{equation}
Now,  eqn.(\ref{umami:eos}) together with eqn.(\ref{20}) represents  implicit expression for $P=P(\rho(T))=P(T)$ and $W=W(\rho(T))=W(T)$. We discuss some following limiting cases.
\paragraph{\underline{\textbf{High energy density: $\rho\rightarrow \infty$}}}:\\

As mentioned earlier when $\rho\rightarrow \infty$, $T$ converges to $\tau$ which gives 
\begin{equation}\label{29}
P=-\sqrt{2|A|\left(1-\frac{T}{\tau}\right)}\rightarrow 0
\end{equation}
and 
\begin{equation}\label{30}
W=\frac{-\sqrt{2|A|\left(1-\frac{T}{\tau}\right)}}{\rho}\rightarrow 0
\end{equation}
So, it implies Umami fluid behaves as dust when $T\rightarrow \tau$ compatible with our previous result.\\

\paragraph{\underline{\textbf{$T=T_{min}$}}}:\\

$T$ attains it's minimum value at $\rho=\rho_1$. At that point the pressure reaches its most negative value $P=-\frac{\sqrt{|A||\omega|}}{2}$ and the EoS parameter $W$ takes value $-\frac{|\omega|}{2}$.\\
 
\paragraph{\underline{\textbf{Low energy density, $\alpha>0 (T\rightarrow \infty)$}}} : \\
In the low energy density limit we find $\rho\approx \sqrt{\frac{|A|}{|\omega|}}\left(\frac{\tau}{T}\right)^\frac{\alpha}{|\omega|}\alpha^{\frac{1+\alpha}{2|\omega|}}$ and then in this limit 
\begin{eqnarray}
  P &\approx & - \sqrt{|A||\omega|}\left(\frac{\tau}{T}\right)^\frac{\alpha}{|\omega|}\alpha^{\frac{1+\alpha}{2|\omega|}}\rightarrow 0\\
 \end{eqnarray}
with constant EoS parameter $W\approx -|\omega|$\\

\paragraph{\underline{\textbf{Low energy density, $\alpha<0,\rho\rightarrow \rho_{min}(T\rightarrow \infty)$}}}:\\
For this case $P\rightarrow -\rho_{min}$ and $W\rightarrow -1$, i.e the di-Sitter value.\\

\section{Observational Analysis}
\label{sec:obs}

The thermodynamic analysis of Sec.~\ref{sec:thermo} splits into two cases. The sign of $\alpha=1-|w|$ sets the case. For $|w|<1$ we have $\alpha>0$. The density falls to zero as the volume grows and the fluid ends as a constant-$w$ fluid. For $1<|w|<2$ we have $\alpha<0$. The density now has a floor $\rho_{\min}$ and the fluid ends in a de Sitter state. The present-epoch stability condition $|A|<|w|\rho_0^2$ is more easily met in the second case. Thermodynamic stability is a physical requirement on the fluid. It is not built into the background fit. We therefore carry both cases into the observational analysis and fit each one separately. The data then decide how well each case works. In Sec.~\ref{sec:results} we compare the quality of each fit with the thermodynamic status of that case. This comparison exposes the tension that is the main result of this paper. Having derived the background expansion rate $E(z)$ in Sec.~\ref{sec:model}, we now confront it with observational data. Before the four datasets described below can be compared with the model, $E(z)$ must first be translated into the distance and time measures that each of them actually constrains; we build these ingredients one at a time.

The starting point for every distance measure is the comoving distance travelled by a photon since redshift $z$,
\begin{equation}
D_C(z) = \frac{c}{H_0}\int_0^{z}\frac{dz'}{E(z')}.
\label{eq:dc}
\end{equation}
For a spatially flat Universe, as assumed throughout this work, the comoving distance coincides with the transverse comoving distance,
\begin{equation}
D_M(z) = D_C(z),
\label{eq:dm}
\end{equation}
from which the two distance measures entering our likelihoods follow directly: the angular diameter distance,
\begin{equation}
D_A(z) = \frac{D_M(z)}{1+z},
\label{eq:da}
\end{equation}
and the luminosity distance,
\begin{equation}
D_L(z) = (1+z)\,D_M(z).
\label{eq:dl}
\end{equation}
Equation~\eqref{eq:dl} is used directly by the supernova likelihood, while Eq.~\eqref{eq:dm} enters the baryon-acoustic-oscillation likelihood together with the sound horizon derived below.

The baryon-acoustic-oscillation likelihood requires the sound horizon at the baryon drag epoch, a quantity set by the physics of the photon-baryon plasma prior to recombination. We compute it using the fitting formulas of \cite{EisensteinHu1998}. The cosmic-microwave-background likelihood requires a different quantity, the sound horizon at photon decoupling, which we compute separately below. The redshift of the drag epoch is
\begin{equation}
z_d = \frac{1291\,(\Om^{\rm eff} h^{2})^{0.251}}{1+0.659\,(\Om^{\rm eff} h^{2})^{0.828}}
\Big[1+b_1(\Ob h^{2})^{b_2}\Big],
\label{eq:zdrag}
\end{equation}
with coefficients
\begin{align}
b_1 &= 0.313\,(\Om^{\rm eff} h^{2})^{-0.419}\Big[1+0.607(\Om^{\rm eff} h^{2})^{0.6748}\Big], \label{eq:b1b2}\\
b_2 &= 0.238\,(\Om^{\rm eff} h^{2})^{0.223},
\notag
\end{align}
and the corresponding comoving sound horizon is
\begin{equation}
r_s(z_d) = \frac{44.5\,\ln\!\left(9.83/\Om^{\rm eff} h^{2}\right)}{\sqrt{1+10\,(\Ob h^{2})^{0.75}}}\ \ \text{Mpc}.
\label{eq:rs}
\end{equation}
The BAO likelihood uses $r_d\equiv r_s(z_d)$ of Eq.~\eqref{eq:rs}. The CMB acoustic scale uses $r_s(z_\ast)$ instead. We compute it by direct integration of the baryon--photon sound speed,
\begin{align}
r_s(z_\ast)&=\int_{z_\ast}^{\infty}\frac{c_s(z)\,dz}{H_0E_{\rm eff}(z)},\notag\\
c_s&=\frac{c}{\sqrt{3\,[1+R_b(z)]}},\quad R_b=\frac{3\,\Omega_b}{4\,\Omega_\gamma}\,\frac{1}{1+z}.
\label{eq:rsstar}
\end{align}
where $E_{\rm eff}$ is given in Eq.~\eqref{eq:Eeff} below and $\Omega_\gamma h^2=2.4728\times10^{-5}$. For the $|w|<1$ best fit we find $r_s(z_d)=150.1$~Mpc and $r_s(z_\ast)=145.1$~Mpc, a difference of $3\%$. The difference grows to $8\%$ for the $|w|>1$ best fit. Using $r_s(z_d)$ in the acoustic scale would bias the CMB constraint, because the measured $\ell_A$ has a precision of $0.03\%$. Equation~\eqref{eq:rs} is a fitting formula. It agrees with a direct integration of the sound speed to within about $4\%$ for all our best fits.
Equations~\eqref{eq:zdrag}--\eqref{eq:rs} are evaluated using the derived $\Om^{\rm eff}$ of Eq.~\eqref{eq:Omeff}, since they are calibrated on standard early-Universe physics rather than on the genuine background dynamics of Eq.~\eqref{eq:friedmann}. For the same reason, the comoving distance to the photon-decoupling redshift $z_\ast=1089.80$, required by the cosmic-microwave-background likelihood, is computed not from Eq.~\eqref{eq:dc} but from an effective flat-$\Lambda$CDM expansion history built from $\Om^{\rm eff}$,
\begin{equation}
E_{\rm eff}(z) = \sqrt{\Om^{\rm eff}(1+z)^{3}+\Or(1+z)^{4}+\left(1-\Om^{\rm eff}-\Or\right)},
\label{eq:Eeff}
\end{equation}
so that the two Planck shift parameters entering the cosmic-microwave-background likelihood are
\begin{equation}
R(z_\ast) \equiv \sqrt{\Om^{\rm eff}}\,\frac{H_0}{c}\,D_M(z_\ast), \qquad
\ell_A(z_\ast) \equiv \pi\,\frac{D_M(z_\ast)}{r_s(z_\ast)},
\label{eq:shiftparams}
\end{equation}
with $D_M(z_\ast)$ obtained from Eq.~\eqref{eq:dm} using $E_{\rm eff}(z)$ of Eq.~\eqref{eq:Eeff} in place of $E(z)$, and $r_s(z_\ast)$ from Eq.~\eqref{eq:rsstar}. By contrast, the genuine background expansion rate of Eq.~\eqref{eq:friedmann} is used without modification throughout the supernova, baryon-acoustic-oscillation, and Hubble-parameter likelihoods, since these probe only $z\lesssim2.3$, well within the regime constrained directly by the data.

\subsection{Datasets}

Table~\ref{tab:datasets} lists the four datasets used to constrain the model. Type Ia supernovae are taken from the Pantheon+SH0ES compilation \cite{Scolnic2022,Brout2022}, which provides 1701 light curves; of these we retain the 1624 objects not used as Cepheid host-galaxy calibrators, so that the resulting constraint is independent of the local distance-ladder calibration of $H_0$. Baryon-acoustic-oscillation measurements are taken from Data Release 2 of the Dark Energy Spectroscopic Instrument \cite{DESI2025}, comprising thirteen measurements of $D_M/r_d$, $D_H/r_d$, and $D_V/r_d$ across seven effective redshifts. Hubble parameter measurements from cosmic chronometers, obtained from the differential ages of passively evolving early-type galaxies, are taken from the compilation summarized by \cite{Moresco2022}. Cosmic-microwave-background information enters through the Planck 2018 distance priors of \cite{Chen2018}, compressed into the shift parameters \cite{WangMukherjee2007} of Eq.~\eqref{eq:shiftparams} together with the physical baryon density $\Ob h^2$.

\begin{table*}[t]
\centering
\small
\caption{Observational datasets used in the statistical analysis.}
\label{tab:datasets}
\begin{tabular}{@{}>{\raggedright\arraybackslash}p{4.0cm}>{\raggedright\arraybackslash}p{5.0cm}>{\centering\arraybackslash}p{2.8cm}>{\raggedright\arraybackslash}p{2.6cm}@{}}
\toprule
Probe & Sample & Redshift range & Reference \\
\midrule
Type Ia Supernovae & Pantheon+SH0ES, 1624 SNe & $0.001<z<2.26$ & \cite{Scolnic2022,Brout2022} \\
Baryon Acoustic Oscillations & DESI DR2, 13 measurements & $0.295<z<2.33$ & \cite{DESI2025} \\
Cosmic chronometers & $H(z)$ compilation, 31 points & $0.07<z<1.97$ & \cite{Moresco2022} \\
CMB distance priors & Planck 2018 shift parameters & $z_\ast=1089.80$ & \cite{Chen2018} \\
\bottomrule
\end{tabular}
\end{table*}

\subsection{Likelihoods}

With the distance and sound-horizon relations of Eqs.~\eqref{eq:dc}--\eqref{eq:shiftparams} in hand, we now construct a Gaussian likelihood, $\mathcal{L}\propto\exp(-\chi^2/2)$, for each of the four datasets in Table~\ref{tab:datasets}. For the supernovae, the theoretical distance modulus follows from Eq.~\eqref{eq:dl},
\begin{equation}
\mu_{\rm th}(z) = 5\log_{10}D_L(z) + 25.
\label{eq:mu}
\end{equation}
The observed apparent magnitude carries an unknown additive offset, degenerate with the supernova absolute magnitude and with $H_0$, which we marginalize over analytically rather than fit as a free parameter. Writing $\Delta F\equiv m_{B,{\rm corr}}-\mu_{\rm th}$ and $C_{\rm SN}$ for the full statistical-plus-systematic covariance matrix of the retained sample,
\begin{equation}
a\equiv\Delta F^{T}C_{\rm SN}^{-1}\Delta F, \qquad
b\equiv\Delta F^{T}C_{\rm SN}^{-1}\mathbf 1, \qquad
d\equiv\mathbf 1^{T}C_{\rm SN}^{-1}\mathbf 1,
\label{eq:snabd}
\end{equation}
the marginalized supernova chi-square is
\begin{equation}
\chi^2_{\rm SN} = a-\frac{b^2}{d}.
\label{eq:chisn}
\end{equation}
For the baryon-acoustic-oscillation data, with $\Delta F_{\rm BAO}$ the difference between the observed and predicted vector of $D/r_d$ ratios and $C_{\rm BAO}$ its covariance matrix,
\begin{equation}
\chi^2_{\rm BAO} = \Delta F_{\rm BAO}^{T}\,C_{\rm BAO}^{-1}\,\Delta F_{\rm BAO}.
\label{eq:chibao}
\end{equation}
For the cosmic-chronometer Hubble measurements, which are statistically independent,
\begin{equation}
\chi^2_{H} = \sum_{i=1}^{31}\frac{\big[H_{\rm th}(z_i)-H_{\rm obs}(z_i)\big]^{2}}{\sigma_{H}^{2}(z_i)}.
\label{eq:chihz}
\end{equation}
For the cosmic-microwave-background distance priors, with $\Delta F_{\rm CMB}=(R,\ell_A,\Ob h^2)_{\rm th}-(R,\ell_A,\Ob h^2)_{\rm obs}$ and $C_{\rm CMB}$ the corresponding covariance matrix,
\begin{equation}
\chi^2_{\rm CMB} = \Delta F_{\rm CMB}^{T}\,C_{\rm CMB}^{-1}\,\Delta F_{\rm CMB}.
\label{eq:chicmb}
\end{equation}
The total likelihood combines all four probes,
\begin{equation}
\chi^2_{\rm tot} = \chi^2_{\rm SN}+\chi^2_{\rm BAO}+\chi^2_{H}+\chi^2_{\rm CMB}.
\label{eq:chitot}
\end{equation}

\subsection{Parameter inference}\label{sec:param_inference}
The model has four free parameters, $\{\Ob,h,w,\tilde A\}$, with flat priors $\Ob\in(0.01,0.1)$, $h\in(0.5,0.85)$, $w\in(-0.999,-0.02)$, and $\tilde A\in(-3,5)$ for the $|w|<1$ case; $\Ofz$ follows from flatness and $\Om^{\rm eff}$ is derived as in Eq.~\eqref{eq:Omeff}, so neither introduces an additional degree of freedom. For the $|w|>1$ case, $w$ is instead confined to the hard bound $w\in(-2.2,-1.001)$ required by Eq.~\eqref{eq:floor} below, and we impose a weak Gaussian prior on $w$ and $A$, centered at $w=-2.0$ and $A=1.2$ with $\sigma=0.05$ on each, added as a penalty term to $\chi^2_{\rm tot}$; $\Ob$ and $h$ keep the same flat priors as the $|w|<1$ case. This prior is an assumption that selects the stable region. To measure its effect we repeat the $|w|>1$ fit without it. This fit keeps the hard bound on $w$ and uses the flat prior $\tilde A\in(-3,5)$. We also find the best fit that satisfies the stability condition $|A|\le|w|\Omega_{f,0}^2$. For this fit we write $A=f\,|w|\Omega_{f,0}^2$ with $0<f\le1$ and minimize $\chi^2_{\rm tot}$ over $\{\Ob,h,w,f\}$ from seven starting points. All starts give the same minimum, which lies at $f=1$. We do not run a chain for this fit. For comparison we fit flat $\Lambda$CDM with $E(z)=\sqrt{\Omega_m(1+z)^3+\Or(1+z)^4+1-\Omega_m-\Or}$ and free parameters $\{\Omega_m,h,\Omega_b\}$ with flat priors $\Omega_m\in(0.1,0.6)$, $h\in(0.5,0.85)$ and $\Omega_b\in(0.01,0.1)$. The sound horizons $r_d$ and $r_s(z_\ast)$ and the CMB distance use $\Omega_m$ in place of $\Om^{\rm eff}$. We first locate the best-fit point with a Nelder--Mead downhill-simplex minimization of $\chi^2_{\rm tot}$ in Eq.~\eqref{eq:chitot} (including the Gaussian-prior penalty for the $|w|>1$ case). We start from several points and keep the lowest minimum. This best-fit point then seeds the affine-invariant ensemble sampler \cite{Goodman2010} \texttt{emcee} \cite{ForemanMackey2013}: for the $|w|<1$ case, 32 walkers are initialized in a small Gaussian ball around it and run for 8000 steps; for the $|w|>1$ case, 32 walkers are run for 10\,000 steps. In both cases the first $25\%$ of each chain is discarded as burn-in and the remainder is thinned by a factor of 5 before extracting the posterior statistics reported in Sec.~\ref{sec:results}. The $\Lambda$CDM and the unrestricted $|w|>1$ fits use the same settings, with 8000 and 10\,000 steps respectively. The integrated autocorrelation times are between 35 and 310 steps. Every chain is more than 25 autocorrelation times long and at least 19 after burn-in.

\section{Results}
\label{sec:results}

We fit the umami model twice. The first fit leaves $w$ free inside the $|w|<1$ case. This is our main fit. The second fit restricts $w$ to the $|w|>1$ case, with a weak Gaussian prior centered at $w=-2.0$, $A=1.2$ ($\sigma=0.05$ on each) as described in Sec.~\ref{sec:param_inference}, motivated by the thermodynamic considerations discussed in Sec.~\ref{sec:thermo}. Everything else is identical between the two fits: same data, same likelihoods, same four free parameters $\{\Ob,h,w,\tilde A\}$. In the tables and figures the label ``$|w|>1$ case'' always means this fit with the prior. We also fit flat $\Lambda$CDM to the same data. Two further $|w|>1$ fits test the role of the prior. They are the fit without the prior and the best fit that satisfies the stability condition. We discuss them after Table~\ref{tab:chi2}.

The $|w|>1$ case has one extra feature not present for $|w|<1$. Its equation of state is only defined above a floor density,
\begin{equation}
\Of^{\rm floor} = \sqrt{|A|\,\frac{|w|-1}{|w|}}.
\label{eq:floor}
\end{equation}
Below this floor the bracket in Eq.~\eqref{umami:eos} turns negative, and the pressure would be complex. We checked that every accepted point in the $|w|>1$ case chain stays above its own floor at all times.

Table~\ref{tab:results} reports the posterior median and $16^{\rm th}$--$84^{\rm th}$ percentile range for each parameter, together with the derived quantities $\Ofz$, $\kappa$, $\Om^{\rm eff}$ and $H_0$, for both cases. The $|w|<1$ case posterior is based on 38\,400 post-burn-in samples, with a mean acceptance fraction of 0.391. The $|w|>1$ case posterior is based on 10\,000 steps with 32 walkers, giving 48\,000 post-burn-in samples, with a mean acceptance fraction of 0.586. The two chains use the same burn-in and thinning.

\begin{table}[tbp]
\centering
\caption{Posterior parameters and derived quantities, $|w|<1$ case versus $|w|>1$ case. The $|w|>1$ case is the fit with the Gaussian prior. Values are posterior medians with $16^{\rm th}$--$84^{\rm th}$ percentile ranges.}
\label{tab:results}
\begin{tabular}{@{}lcc@{}}
\toprule
Quantity & $|w|<1$ case & $|w|>1$ case \\
\midrule
$\Ob$ & $0.0502\pm0.0013$ & $0.0421\pm0.0005$ \\
$h$ & $0.6673^{+0.0085}_{-0.0084}$ & $0.7287\pm0.0040$ \\
$w$ & $-0.7340^{+0.0475}_{-0.0563}$ & $-1.9878^{+0.0468}_{-0.0473}$ \\
$\tilde A$ & $-0.5701^{+0.0606}_{-0.0611}$ & $-0.1165^{+0.0050}_{-0.0053}$ \\
\midrule
\multicolumn{3}{@{}l}{\textit{Derived quantities}} \\
$\Ofz$ & 0.9498 & 0.9579 \\
$\kappa$ & 0.2693 & 0.2773 \\
$\Om^{\rm eff}$ & 0.3194 & 0.3194 \\
$H_0$ [km\,s$^{-1}$\,Mpc$^{-1}$] & 66.8 & 72.9 \\
\bottomrule
\end{tabular}
\end{table}

\begin{figure}[tbp]
\centering
\begin{subfigure}[t]{0.49\textwidth}
\centering
\includegraphics[width=\textwidth]{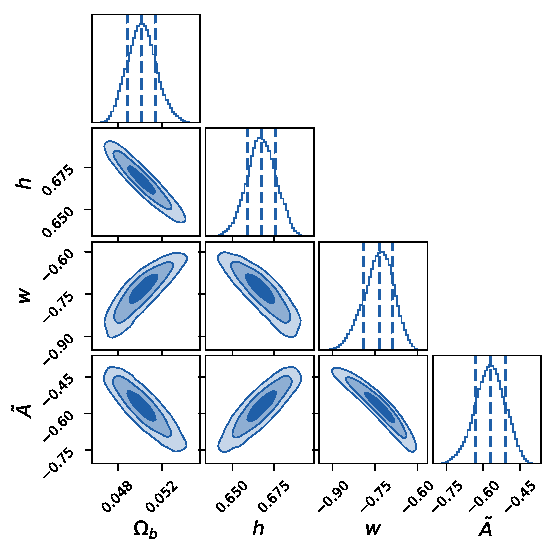}
\caption{$|w|<1$ case.}
\end{subfigure}
\hfill
\begin{subfigure}[t]{0.49\textwidth}
\centering
\includegraphics[width=\textwidth]{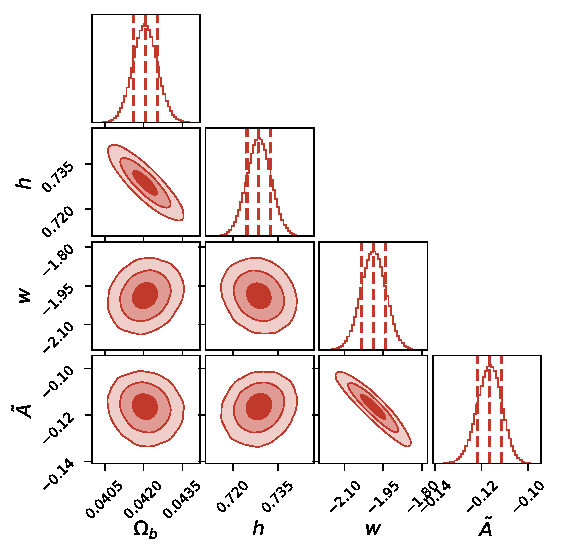}
\caption{$|w|>1$ case.}
\end{subfigure}
\caption{Posterior distributions for $\{\Ob,h,w,\tilde A\}$. Shading from dark to light marks the $1\sigma$, $2\sigma$ and $3\sigma$ credible regions. Dashed lines mark the 16th, 50th and 84th percentiles.}
\label{fig:corner}
\end{figure}

\noindent All four parameters are tightly constrained in the $|w|<1$ case, despite the mutual degeneracy between them (Fig.~\ref{fig:corner}). In the $|w|>1$ case the data constrain $\Ob$ and $h$. The posterior of $w$ is set mainly by the Gaussian prior. Its width, $0.047$, is close to the prior width of $0.05$. The data constrain $A$ more tightly than the prior does. They shift $A$ from $1.2$ to $1.308\pm0.016$. In the $|w|<1$ case the equation-of-state parameters place the fluid away from the pure Chaplygin-gas ($w\to-1$) and pure dust ($w\to0$) limits. Both terms of Eq.~\eqref{umami:eos} are therefore needed. The derived $\Om^{\rm eff}=0.319$ is consistent with independent late-time determinations of the total matter density. The value $H_0=66.8$~km~s$^{-1}$~Mpc$^{-1}$ agrees with the Planck value and is lower than local distance-ladder values. The $|w|>1$ case gives the same $\Om^{\rm eff}=0.319$ but a higher $H_0=72.9$~km~s$^{-1}$~Mpc$^{-1}$. For comparison, flat $\Lambda$CDM gives $\Omega_m=0.3191\pm0.0010$, $h=0.6645\pm0.0035$ and $\Omega_b=0.0505\pm0.0005$. Inserting the best-fit values into the stability condition of Sec.~3.3, the $|w|<1$ case gives $|A|\simeq3.72$, well above $|w|\Omega_{f,0}^2\simeq0.66$, so the present epoch lies on the thermodynamically unstable branch. The $|w|>1$ case gives $|A|\simeq1.31<|w|\Omega_{f,0}^2\simeq1.83$, so it lies on the stable branch. Every sample of the $|w|>1$ chain satisfies the condition.

\begin{table}[tbp]
\centering
\caption{$\chi^2/N$ by dataset at the best-fit point of each model. The $|w|>1$ column is the fit with the Gaussian prior and excludes the penalty term.}
\label{tab:chi2}
\begin{tabular}{@{}lccc@{}}
\toprule
Dataset & $|w|<1$ & $|w|>1$ & $\Lambda$CDM \\
\midrule
Pantheon+ SNe Ia & 0.897 & 0.922 & 0.898 \\
DESI DR2 BAO & 0.771 & 3.131 & 1.258 \\
Cosmic chronometers $H(z)$ & 0.517 & 0.503 & 0.510 \\
Planck 2018 CMB priors & 0.051 & 0.042 & 0.045 \\
\midrule
\textbf{Total} & \textbf{0.888} & \textbf{0.930} & \textbf{0.892} \\
\bottomrule
\end{tabular}
\end{table}

\noindent The $|w|<1$ case fits every probe at or below $\chi^2/N\simeq1$. The supernova and chronometer fits of the $|w|>1$ case are of similar quality. Its CMB fit is excellent, with $\chi^2/N=0.04$. The BAO data show the clearest cost, with $\chi^2/N=3.1$ against $0.77$. The total is $\chi^2=1554.4$ against $1483.0$. About $41$ units of the difference come from the supernovae and $31$ from BAO.

Both umami cases reach the same $\Om^{\rm eff}=0.319$. The CMB and BAO data fix this value through the sound horizon. The two cases differ in how they shape the late-time expansion. The prior holds $w$ near $-2$. This keeps the fluid on the stable branch and changes the shape of $E(z)$ at low redshift. The best fit has $h=0.729$ against $0.667$ in the $|w|<1$ case. The supernovae and BAO data disfavour this combination.

Without the prior the $|w|>1$ fit moves to $w=-1.023$ and falls on the unstable branch, with $\chi^2=1497.9$. The best fit that satisfies the stability condition has $\chi^2=1538.0$, which is $55$ units above the $|w|<1$ best fit. We adopt the $|w|<1$ case as the fiducial fit because it has the lowest $\chi^2$.

\noindent In both cases the one-dimensional marginals are unimodal and approximately Gaussian, and the nested $1\sigma$--$3\sigma$ contours are smooth and closed, indicating a well-converged posterior with no secondary modes. The two-dimensional panels reveal significant correlations in both cases: $\Ob$ and $h$ are anti-correlated, as expected from their joint role in setting the sound horizon, while $w$ and $\tilde A$ are strongly anti-correlated because both parameters jointly control the redshift at which the fluid transitions from matter-like to dark-energy-like behaviour. No contour touches the edge of its flat prior range in either case. In the $|w|>1$ case the posterior of $w$ is set mainly by the Gaussian prior, which is part of the model reported here.

\begin{figure}[tbp]
\centering
\begin{subfigure}[t]{0.49\textwidth}
\centering
\includegraphics[width=\textwidth]{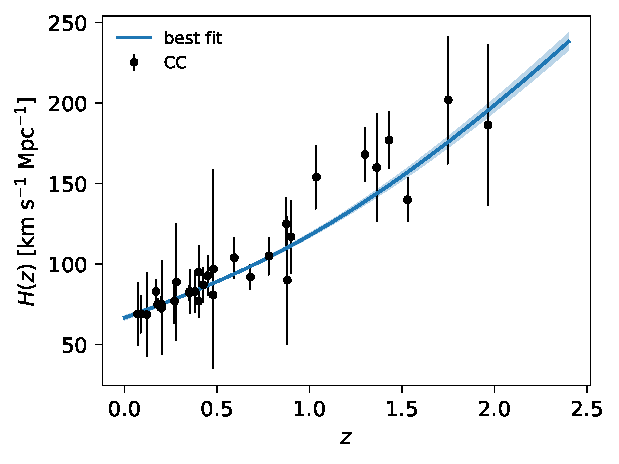}
\caption{$|w|<1$ case.}
\end{subfigure}
\hfill
\begin{subfigure}[t]{0.49\textwidth}
\centering
\includegraphics[width=\textwidth]{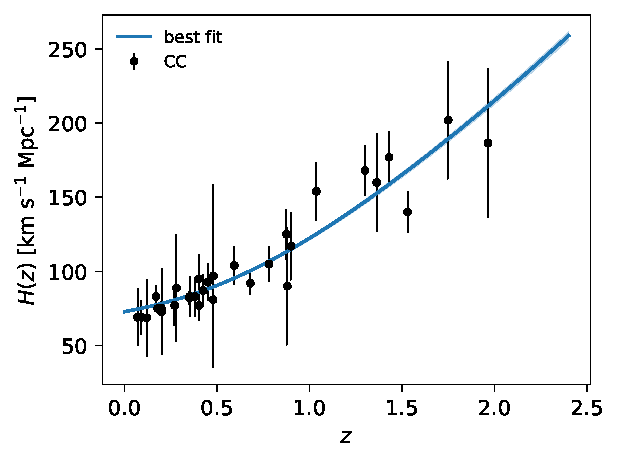}
\caption{$|w|>1$ case.}
\end{subfigure}
\caption{Cosmic-chronometer $H(z)$ measurements compared with the best fit of each case.}
\label{fig:hzfit}
\end{figure}

\noindent The best-fit curve passes through the bulk of the cosmic-chronometer data within their reported uncertainties across the full sampled redshift range, with no systematic trend in the residuals at low or high redshift. The larger error bars at $z\simeq0.3$--$0.5$, where several independent chronometer measurements overlap, are accommodated without difficulty, and the model's slightly higher curvature relative to a naive linear extrapolation reflects the fluid's transition from matter-like to dark-energy-like behaviour captured by Eq.~\eqref{eq:friedmann}. The agreement extends smoothly to the highest-redshift points near $z\simeq2$, where the reported uncertainties are largest. This dataset is fit comparably well by both cases (Table~\ref{tab:chi2}), and the two best-fit curves in Fig.~\ref{fig:hzfit} are correspondingly close to one another.

\begin{figure}[tbp]
\centering
\begin{subfigure}[t]{0.49\textwidth}
\centering
\includegraphics[width=\textwidth]{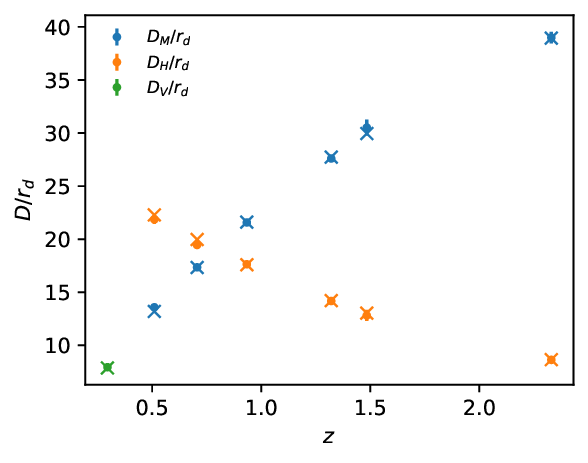}
\caption{$|w|<1$ case.}
\end{subfigure}
\hfill
\begin{subfigure}[t]{0.49\textwidth}
\centering
\includegraphics[width=\textwidth]{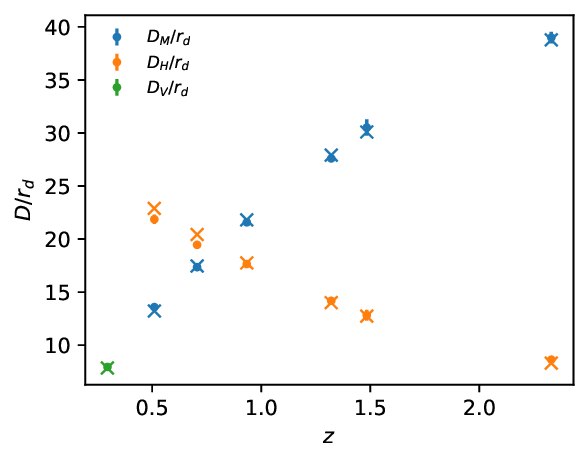}
\caption{$|w|>1$ case.}
\end{subfigure}
\caption{DESI DR2 BAO distance-ratio measurements compared with the best fit of each case.}
\label{fig:baofit}
\end{figure}

\noindent For the $|w|<1$ case, the predicted $D/r_d$ values track the DESI DR2 measurements closely across the full redshift range from $z=0.295$ to $z=2.33$, including both the transverse ($D_M/r_d$) and radial ($D_H/r_d$) distance ratios. The largest residual is the $D_M/r_d$ point at $z=0.51$, at $2.3\sigma$. Here and below the residuals are in units of the diagonal errors. The overall $\chi^2/N=0.77$ in Table~\ref{tab:chi2} shows that no single point drives the fit. The agreement at the lowest and highest redshift bins provides an independent check on the derived $\Om^{\rm eff}$ of Eq.~\eqref{eq:Omeff}. The $|w|>1$ case fits the same data with $\chi^2/N=3.1$. Its largest residuals are the $D_H/r_d$ points at $z=0.51$, $z=0.71$ and $z=2.33$, at $2.4\sigma$, $2.9\sigma$ and $-3.3\sigma$.

\paragraph{Model comparison.}
We compare the models with the Akaike and Bayesian information criteria \cite{Akaike1974,Schwarz1978,Liddle2007}. They are defined as
\begin{equation}
{\rm AIC}=\chi^2_{\min}+2k, \qquad {\rm BIC}=\chi^2_{\min}+k\ln N,
\label{eq:aicbic}
\end{equation}
where $k$ is the number of free parameters and $N$ is the number of data points. We use $N=1671$. The umami cases have $k=4$ and $\Lambda$CDM has $k=3$. For the $|w|>1$ fit with the Gaussian prior we use the data part of $\chi^2$ only. Table~\ref{tab:aicbic} lists the results. The $|w|<1$ case has $\Delta{\rm AIC}=-4.8$ and $\Delta{\rm BIC}=+0.6$ relative to $\Lambda$CDM. The AIC mildly prefers the umami fluid. The BIC does not. The two models are statistically comparable \cite{Liddle2007}. The small gain comes almost entirely from the BAO data, with $\chi^2=10.0$ against $16.4$. The $|w|>1$ case with the prior is strongly disfavoured, with $\Delta{\rm AIC}=+66.5$ and $\Delta{\rm BIC}=+71.9$. Part of this cost comes from the prior. Without it the fit moves to $w\simeq-1$ and has $\Delta{\rm AIC}=+10.0$. This point is unstable. The best fit that satisfies the stability condition has $\Delta{\rm AIC}=+50.1$ and $\Delta{\rm BIC}=+55.6$. The preference against stable fits therefore does not depend on the prior. The information criteria do not change the thermodynamic status of the $|w|<1$ case, which is still unstable at the present epoch.

\begin{table}[tbp]
\centering
\caption{Information criteria. Here $\Delta$ is the difference from $\Lambda$CDM. The label ``prior'' marks the fit with the Gaussian prior on $w$ and $A$. The label ``free'' marks the fit without it. The label ``stable'' marks the best fit that satisfies $|A|\le|w|\Omega_{f,0}^2$. This fit has no chain. For ``prior'' the $\chi^2_{\min}$ excludes the penalty.}
\label{tab:aicbic}
\small
\setlength{\tabcolsep}{3.5pt}
\begin{tabular}{@{}lcccccc@{}}
\toprule
Model & $k$ & $\chi^2_{\min}$ & AIC & BIC & $\Delta$AIC & $\Delta$BIC \\
\midrule
$\Lambda$CDM & 3 & 1489.9 & 1495.9 & 1512.1 & 0 & 0 \\
$|w|<1$ & 4 & 1483.0 & 1491.0 & 1512.7 & $-4.8$ & $+0.6$ \\
$|w|>1$ (prior) & 4 & 1554.4 & 1562.4 & 1584.1 & $+66.5$ & $+71.9$ \\
$|w|>1$ (free) & 4 & 1497.9 & 1505.9 & 1527.5 & $+10.0$ & $+15.4$ \\
$|w|>1$ (stable) & 4 & 1538.0 & 1546.0 & 1567.7 & $+50.1$ & $+55.6$ \\
\bottomrule
\end{tabular}
\end{table}

\section{Perturbative Stability Analysis}
\label{sec:perturbations}

A fluid proposed to unify dark matter and dark energy must satisfy a requirement beyond reproducing the correct background expansion history: it must be capable of gravitationally clustering, in the manner of cold dark matter, at early and intermediate times, while remaining smooth enough at late times to drive accelerated expansion without disrupting the observed large-scale structure. Whether the umami fluid satisfies this requirement cannot be established from the background analysis of Secs.~\ref{sec:model}--\ref{sec:results} alone, since the Friedmann equation is insensitive to the local response of the fluid to density perturbations. We therefore examine the fluid's behaviour under linear perturbations, proceeding in two steps: first, we determine the sign and magnitude of the fluid's sound speed across cosmic history, since this single quantity controls whether density perturbations propagate as stable oscillations or grow without bound; second, we solve the coupled evolution of density perturbations in the umami fluid and in baryons, since the growth of structure depends on the perturbations of the total gravitating matter rather than on the sign of the sound speed alone.

For a barotropic fluid, the response of pressure to a density perturbation is governed by the adiabatic sound speed $v_s^2\equiv dP/d\rho$ \cite{Hu1998}. If $v_s^2$ is positive, a local compression is met by a restoring pressure gradient, so that density perturbations oscillate acoustically, as in an ordinary fluid; if $v_s^2$ is negative, the same term acts as a driving rather than a restoring force, and any small perturbation grows exponentially rather than oscillating -- a tachyonic instability. Establishing the sign of $v_s^2$ for the umami fluid is therefore the first and most direct diagnostic of whether it can support physically sensible density perturbations. Differentiating Eq.~\eqref{umami:eos}, with $B\equiv1/|w|$ and $C\equiv1/|A|$,
\begin{equation}
v_s^2(\rho) = \frac{C\rho^2-B}{(B+C\rho^2)^2},
\label{eq:vs2}
\end{equation}
which tends to $v_s^2\to|A|/\rho^2\to0^+$ as $\rho\to\infty$, reproducing the always-non-negative sound speed of the original Chaplygin gas, but tends to $v_s^2\to-1/B=-|w|$ as $\rho\to0$ -- strictly negative for any $w\neq0$. This is a direct and unavoidable consequence of the fluid's low-density limit being a genuine constant-$w$ fluid rather than a pure cosmological constant: the same feature of Eq.~\eqref{umami:eos} that gives the model its extra degree of freedom is also responsible for a negative sound speed at late times. The critical density at which $v_s^2$ changes sign is $\Of^{\rm crit}=\sqrt{|A|/|w|}$. This is why we also tested the $|w|>1$ case in Sec.~\ref{sec:results}: moving $w$ and $A$ into this region changes $\Of^{\rm crit}$, and we check below whether it moves the instability out of the observable range.

\begin{figure}[tbp]
\centering
\begin{subfigure}[t]{0.49\textwidth}
\centering
\includegraphics[width=\textwidth]{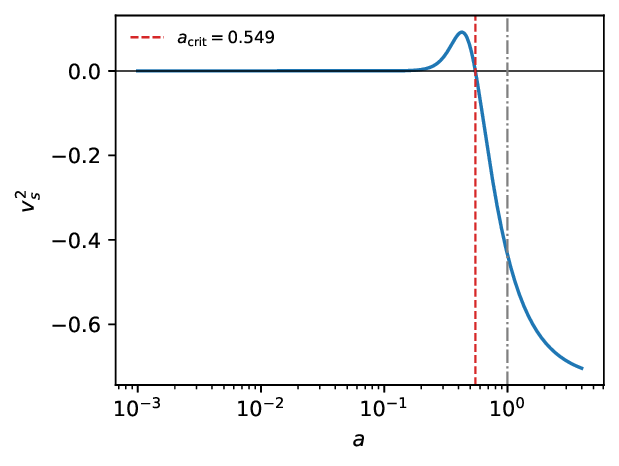}
\caption{$|w|<1$ case, $a_{\rm crit}=0.549$.}
\end{subfigure}
\hfill
\begin{subfigure}[t]{0.49\textwidth}
\centering
\includegraphics[width=\textwidth]{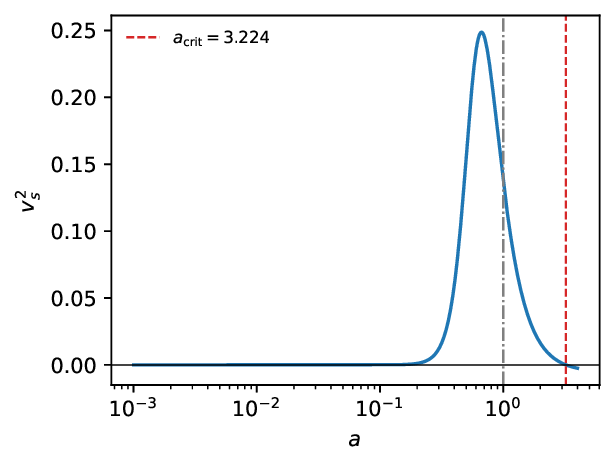}
\caption{$|w|>1$ case, $a_{\rm crit}=3.224$ (in the future). The dash-dotted line marks $a=1$ (today).}
\end{subfigure}
\caption{Sound speed squared of the umami fluid as a function of scale factor.}
\label{fig:vs2}
\end{figure}

For the $|w|<1$ case, $\Of^{\rm crit}=2.253$, to be compared with $\Ofz=0.950$, so the transition from a stable to an unstable sound speed occurred in the past. Solving $\Of(a_{\rm crit})=\Of^{\rm crit}$ using Eq.~\eqref{eq:implicit} gives
\begin{equation}
a_{\rm crit}=0.549, \qquad z_{\rm crit}=0.822,
\label{eq:acrit}
\end{equation}
with $v_s^2(a=1)=-0.435$ today. For the $|w|>1$ case, $\Of^{\rm crit}=0.8106$, to be compared with $\Ofz=0.9579$. Here $\Of^{\rm crit}$ sits \emph{below} today's density rather than above it, so the transition has not yet happened: it lies in the future,
\begin{equation}
a_{\rm crit}=3.224, \qquad z_{\rm crit}=-0.690,
\label{eq:acritphantom}
\end{equation}
with $v_s^2(a=1)=+0.137$ today. So unlike the $|w|<1$ case, the $|w|>1$ case is not unstable today: its sound speed is positive throughout the entire observed history, $z\geq0$, and only turns negative after today, at $z_{\rm crit}=-0.690$. Moving $w$ and $A$ into this deep region of the $|w|>1$ case pushes the instability out of the observed past and present entirely, at the cost in fit quality reported in Tables~\ref{tab:chi2} and~\ref{tab:aicbic}. Figure~\ref{fig:vs2} shows $v_s^2(a)$ for both cases side by side.

Knowing that $v_s^2$ turns negative at $a_{\rm crit}$ establishes that a local instability exists, but does not by itself determine how a realistic density perturbation evolves in the presence of gravity and of the baryons that share the Universe with the umami fluid. We therefore solve the linear evolution of the density contrasts $\delta_b$ (baryons) and $\delta_f$ (umami fluid) self-consistently, for both cases, following the Newtonian perturbative approach used by \cite{Fabris2002} for the original Chaplygin gas, generalized here to two independently conserved fluids coupled only through their common gravitational potential (no direct interaction or energy exchange between them). Linearizing the continuity, Euler, and Poisson equations for each species and writing the result in e-folding time $N\equiv\ln a$,
\begin{align}
&\delta_b'' + \Big[2+\frac{H'}{H}\Big]\delta_b'
\notag\\
&\;- \frac{3}{2}\Big(\frac{H_0}{H}\Big)^{2}\Big[\Omega_b(a)\delta_b+\Of(a)\delta_f\Big] = 0, \label{eq:pertb}\\[4pt]
&\delta_f'' + \Big[2+\frac{H'}{H}\Big]\delta_f' + \Big(\frac{k}{aH}\Big)^{2}v_s^{2}(a)\,\delta_f
\notag\\
&\;- \frac{3}{2}\Big(\frac{H_0}{H}\Big)^{2}\Big[\Omega_b(a)\delta_b+\Of(a)\delta_f\Big] = 0, \label{eq:pertf}
\end{align}
where $k$ is the comoving wavenumber and the two species are coupled through the total density perturbation sourcing the gravitational potential. The mechanism by which the instability identified in Eq.~\eqref{eq:vs2} propagates into these equations is now explicit: for $a<a_{\rm crit}$, the term $(k/aH)^2v_s^2\delta_f$ in Eq.~\eqref{eq:pertf} is restoring, and $\delta_f$ oscillates with bounded amplitude around the overall growing mode sourced by gravity; for $a>a_{\rm crit}$, $v_s^2<0$ and this same term becomes a driving, tachyonic term, so that $\delta_f$ grows exponentially rather than oscillating. Because the destabilizing term scales as $k^2$, the effect is strongly scale-dependent, with smaller-scale (larger-$k$) perturbations growing far more rapidly than large-scale ones. No closed-form solution to Eqs.~\eqref{eq:pertb}--\eqref{eq:pertf} exists, since $\Of(a)$ and $v_s^2(a)$ are themselves only implicitly defined through Eq.~\eqref{eq:implicit}; we therefore integrate the system numerically with an adaptive implicit Runge--Kutta method, starting from growing-mode initial conditions deep in the matter-dominated era.

\begin{figure}[tbp]
\centering
\begin{subfigure}[t]{0.49\textwidth}
\centering
\includegraphics[width=\textwidth]{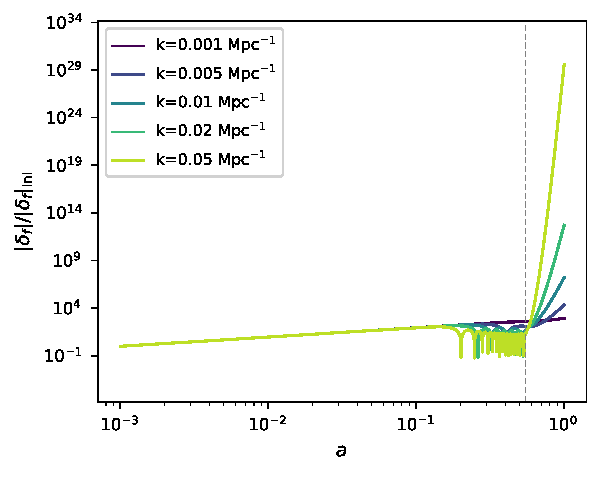}
\caption{$|w|<1$ case.}
\end{subfigure}
\hfill
\begin{subfigure}[t]{0.49\textwidth}
\centering
\includegraphics[width=\textwidth]{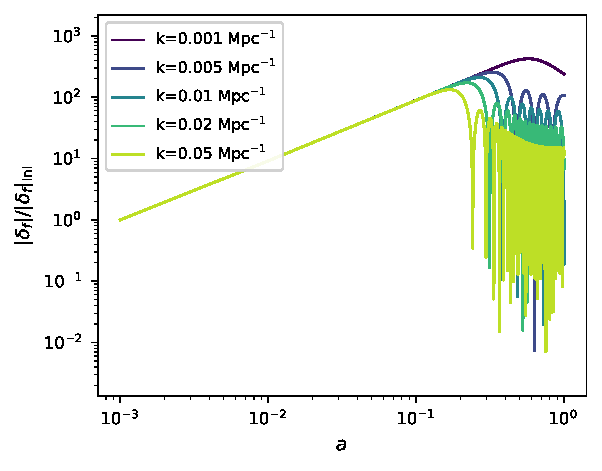}
\caption{$|w|>1$ case. No runaway growth occurs by $a=1$, since $a_{\rm crit}>1$ for this case.}
\end{subfigure}
\caption{Evolution of the umami density contrast for several comoving wavenumbers. The dashed line in panel (a) marks $a_{\rm crit}$.}
\label{fig:growth}
\end{figure}

The full numerical solution of Eqs.~\eqref{eq:pertb}--\eqref{eq:pertf} confirms the qualitative picture above, but it is useful to have an independent, analytic estimate of the growth rate as well, both as a cross-check on the numerical integration and to make the dependence on wavenumber $k$ explicit. We obtain this using the Wentzel--Kramers--Brillouin (WKB) approximation, a standard technique for equations whose coefficients vary slowly compared to the solution's own rate of change. For $a>a_{\rm crit}$ and at sufficiently large $k$, the destabilizing term $(k/aH)^2v_s^2\delta_f$ in Eq.~\eqref{eq:pertf} dominates over the Hubble friction and gravitational source terms, so that the equation reduces approximately to
\begin{equation}
\delta_f'' - \left(\frac{k}{aH}\right)^{2}|v_s^2(a)|\,\delta_f \approx 0,
\label{eq:wkbapprox}
\end{equation}
where we have written $v_s^2=-|v_s^2|$ since $v_s^2<0$ in this regime. Because the coefficient $(k/aH)^2|v_s^2(a)|$ varies slowly in $N$ compared to the rate at which $\delta_f$ itself grows, Eq.~\eqref{eq:wkbapprox} can be solved locally with the WKB ansatz $\delta_f\equiv\exp[S(N)]$. Substituting this ansatz into Eq.~\eqref{eq:wkbapprox} gives
\begin{equation}
\left(S'\right)^{2} + S'' = \left(\frac{k}{aH}\right)^{2}|v_s^2(a)|,
\label{eq:wkbsub}
\end{equation}
and neglecting $S''$ relative to $(S')^2$ -- the adiabatic condition that defines the WKB approximation, valid precisely when the coefficient on the right-hand side varies slowly enough that $S$ itself does not accelerate appreciably over one e-fold -- reduces Eq.~\eqref{eq:wkbsub} to the algebraic relation $S'\approx\pm(k/aH)\sqrt{|v_s^2|}$, which integrates directly to $S(N)=\pm\int(k/aH)\sqrt{|v_s^2|}\,dN$, i.e.\ $\delta_f\propto\exp\!\big[\pm\!\int(k/aH)\sqrt{|v_s^2|}\,dN\big]$. Retaining the growing branch and integrating from $N_{\rm crit}\equiv\ln a_{\rm crit}$, where the approximation becomes valid, up to today ($N=0$) gives the growth factor accumulated once $a$ exceeds $a_{\rm crit}$,
\begin{equation}
\delta_f(a{=}1) \sim \delta_f(a_{\rm crit})\exp\!\left[k\int_{N_{\rm crit}}^{0}\frac{\sqrt{|v_s^2(a)|}}{aH(a)}\,dN\right],
\label{eq:wkb}
\end{equation}
which gives a coefficient of $1.32\times10^3$~Mpc for the $|w|<1$ case. The predicted growth factor after $a_{\rm crit}$ rises from about $4$ at $k=10^{-3}$~Mpc$^{-1}$ to about $4\times10^{28}$ at $k=5\times10^{-2}$~Mpc$^{-1}$. The full numerical solution of Eqs.~\eqref{eq:pertb}--\eqref{eq:pertf}, shown in Fig.~\ref{fig:growth}(a), gives $2$ and $1.6\times10^{28}$. The exponents agree to better than $2\%$ at large $k$.

The same equations apply to the $|w|>1$ case, with its own $a_{\rm crit}$ and $v_s^2(a)$, but here the WKB growth estimate of Eq.~\eqref{eq:wkb} does not apply at all by today: since $a_{\rm crit}=3.224>1$ for this case, the destabilizing regime $a>a_{\rm crit}$ is never reached within the observed history, and $\delta_f$ remains in the oscillatory regime throughout $0<a\leq1$. Figure~\ref{fig:growth}(b) shows this directly: $\delta_f$ oscillates with bounded amplitude across the full range plotted, with no sign of the runaway growth seen in panel (a). Moving $w$ and $A$ into this deep region of the $|w|>1$ case therefore does more than delay the instability: it removes the instability entirely from the range $z\geq0$ actually probed by the data used in Sec.~\ref{sec:obs}. The trade-off, set out in Tables~\ref{tab:chi2} and~\ref{tab:aicbic}, is that this region of parameter space fits the data worse than the fiducial $|w|<1$ case.

\noindent This behaviour identifies a tension within the unification scenario studied here: the same low-density equation-of-state limit that allows the umami fluid to act as a dark-energy component with $w\neq-1$ also produces a negative adiabatic sound speed and a consequent small-scale instability once that regime is reached, in line with the general obstruction to unified dark-sector fluids identified by \cite{Sandvik2004}. The fiducial $|w|<1$ case is already unstable today. Moving $w$ and $A$ into the deep region of the $|w|>1$ case explored here, motivated by the thermodynamic considerations of Sec.~\ref{sec:thermo}, removes this problem over the full observed range $z\geq0$, since the sound speed stays positive until $z_{\rm crit}=-0.690$, strictly in the future -- but only at a cost in the fit of at least $\Delta\chi^2\simeq55$, as shown in Table~\ref{tab:aicbic}. This is a separate finding from the dynamical-systems stability of the background evolution reported by \cite{Huang2021}, which concerns the stability of fixed points of the autonomous background system rather than the linear stability of density perturbations at finite $k$ studied here; a model may be stable in the former sense while exhibiting the instability found in this section. Two developments in the literature are directly relevant to whether the $|w|<1$ case's instability ultimately precludes viable structure formation: the nonlinear treatment of \cite{Abdullah2022}, which shows that collapse can proceed above a model-dependent Jeans scale even where linear theory predicts runaway growth, and the non-adiabatic perturbation schemes of \cite{Reis2003} and \cite{Hashim2025}, which decouple the perturbation sound speed from the adiabatic value of Eq.~\eqref{eq:vs2} and have been shown to remove analogous instabilities in related Chaplygin-type models.



\section{Conclusion}
\label{sec:conclusion}

In this work we have studied a spatially flat FRW universe filled with baryons, radiation, and a single unified dark fluid, the umami Chaplygin gas (UCG), with no separate dark-matter component. The UCG equation of state interpolates between pressureless dust at high density and a barotropic fluid with constant equation-of-state parameter $w$ at low density, so that a single fluid can play the role of dark matter at early times and of dark energy at late times. Because the fluid's continuity equation admits only an implicit solution, we reduced the background evolution to a one-dimensional root-finding problem for the monotonic function $F(\Omega_f)$ of Eq.~\eqref{eq:Fdef}, giving a fully specified four-parameter background model $\{\Omega_b, h, w, \tilde{A}\}$.

We first examined the thermodynamic behaviour of the fluid following the formalism of Santos et al. The condition $(\partial P/\partial V)_T = 0$ holds identically, since the pressure depends on temperature alone. However, $(\partial P/\partial V)_S < 0$ and $C_V > 0$ are satisfied only for $\rho > \rho_1 = \sqrt{|A/\omega|}$. Unlike the generalized and modified Chaplygin gases, which are thermodynamically stable throughout their evolution and cool monotonically, the UCG reaches a minimum temperature at $\rho = \rho_1$ and reheats thereafter, so its thermodynamic stability is conditional rather than global. A viable low-density regime further restricts $|\omega| < 2$. However, whether the present epoch lies in the stable regime completely depends on the value of $A$ and $\omega$. More precisely if the condition $\rho_0^2>|\frac{A}{\omega}|$ is satisfied for the present density $\rho_0$, the current phase will definitely lie in the stable regime and so also the past. With the present density fixed by flatness, $\rho_0 = 1 - \Omega_b - \Omega_r \simeq 0.95$, from the best fits we find that $\rho_0^2>|\frac{A}{\omega}|$ fails for $|w|<1$. The present epoch then lies on the unstable reheating branch. For $|w|>1$ the condition holds in the fit with the Gaussian prior. There the present epoch lies on the stable cooling branch and the fluid approaches a de Sitter end state resembling $\Lambda$CDM. The condition fails for $|w|>1$ in the fit without the prior.

Motivated by this, we confronted both cases with Pantheon+ supernovae, DESI DR2 BAO, cosmic-chronometer $H(z)$ data and Planck 2018 distance priors. We also fitted flat $\Lambda$CDM to the same data. The $|w|<1$ case describes the background expansion well, with $w = -0.734^{+0.048}_{-0.056}$, $\Omega_m^{\rm eff} = 0.319$, $H_0=66.8$ km s$^{-1}$ Mpc$^{-1}$ and a total $\chi^2/N = 0.888$. The data require both terms of the equation of state, placing the fluid away from the pure Chaplygin-gas and pure-dust limits. This case is statistically comparable to $\Lambda$CDM, with $\Delta{\rm AIC}=-4.8$ and $\Delta{\rm BIC}=+0.6$. The thermodynamically favoured $|w|>1$ case with the Gaussian prior raises $H_0$ to $72.9$ km s$^{-1}$ Mpc$^{-1}$ but fits worse, with $\Delta{\rm AIC}=+66.5$ relative to $\Lambda$CDM and a BAO $\chi^2/N$ of $3.1$. Part of this cost comes from our Gaussian prior. Without it the fit moves to $w\simeq-1$ and falls on the unstable branch. The best fit that satisfies the stability condition exactly has $\Delta{\rm AIC}=+50.1$.

At the perturbative level, the adiabatic sound speed of the UCG is positive at high density but tends to $-|w|$ at low density, changing sign at $\Omega_f^{\rm crit} = \sqrt{|A|/|w|}$. This is exactly the critical density $\rho_1$ found in the thermodynamic analysis, so the thermodynamic and perturbative instabilities are two manifestations of the same feature of the equation of state. In the observationally preferred $|w|<1$ case the transition occurred at $z_{\rm crit} = 0.822$, and small-scale density perturbations have since grown exponentially, by nearly thirty orders of magnitude at $k \sim 0.05$ Mpc$^{-1}$, as confirmed both numerically and through a WKB estimate. In the $|w|>1$ case the transition is pushed into the future ($z_{\rm crit} = -0.690$), and perturbations remain bounded throughout the observed history, in agreement with the thermodynamic prediction.

Taken together, our results expose a tension within the umami unification scenario. The region of parameter space favoured by the background data is both thermodynamically and perturbatively unstable at the present epoch. The region that is stable over the observed history fits the data worse. The best stable fit has $\Delta\chi^2\simeq55$ relative to the $|w|<1$ fit. It has $\Delta{\rm AIC}\simeq+50$ relative to $\Lambda$CDM. Whether this tension can be resolved remains an open question, leading to several directions for future work. A nonlinear study using the spherical-collapse approach of \cite{Abdullah2022} can tell us whether structures such as galaxies and clusters can still form despite the linear instability. Since the instability arises because the sound speed of the fluid becomes negative, allowing non-adiabatic perturbations, for which the effective sound speed need not follow Eq.~\eqref{eq:vs2}, may remove it, as has been shown for other Chaplygin-type models \cite{Reis2003,Hashim2025}. The stable $|w|>1$ case gives $H_0=72.9$ km s$^{-1}$ Mpc$^{-1}$ and the $|w|<1$ case gives $66.8$. Following \cite{Frion2024}, a study that includes the local distance-ladder calibration can test whether the stable branch eases the Hubble tension. Finally, nested sampling would give the Bayesian evidence and allow a direct evidence-based comparison with $\Lambda$CDM.

\end{document}